\documentclass[journal,article,submit,moreauthors]{mdpi} 
\UseRawInputEncoding
\firstpage{1} 
\hreflink{https://doi.org/} 

\usepackage{wrapfig}
\usepackage[pscoord]{eso-pic}
\usepackage[fulladjust]{marginnote}
\usepackage{tikz}
\usepackage[many]{tcolorbox}
\tcbuselibrary{skins, breakable, theorems}

\usepackage{varwidth}
\usepackage{extarrows}
\usepackage{tikz}
\usepackage{tabu}
\usetikzlibrary{arrows}
\usepackage{tcolorbox}
\tcbuselibrary{skins, breakable, theorems}
\usepackage{fancybox}
\usepackage{amsthm}

\reversemarginpar

\definecolor{Gray}{gray}{.25}
\definecolor{lightgray1}{gray}{0.95}
\definecolor{lightgray2}{gray}{0.6}
\definecolor{lightgray3}{gray}{0.8}

\Title{\boldmath Long-range potential scattering: Converting long-range potential to
short-range potential by tortoise coordinate} 

\Author{Wen-Du Li $^{1,2,3,\dagger}$ , Wu-Sheng Dai $^{2,\ddagger}$}

\address{%
$^{1}$ \quad College of Physics and Materials Science, Tianjin Normal University, Tianjin 300387, PR China\\
$^{2}$ \quad Department of Physics, Tianjin University, Tianjin 300350, PR China\\
$^{3}$ \quad Theoretical Physics Division, Chern Institute of Mathematics, Nankai University, Tianjin, 300071, PR China
}

\firstnote{liwendu@tjnu.edu.cn.} 
\secondnote{daiwusheng@tju.edu.cn.}

\abstract{Inspired by general relativity, we suggest an approach for long-range
potential scattering. In scattering theory, there is a general theory for
short-range potential scattering, but there is no general theory\ for
long-range potential scattering. This is because the scattering boundary
conditions for all short-range potentials are the same, but for different
long-range potentials are different. In this paper, by introducing tortoise
coordinates, we convert long-range potential scattering to short-range
potential scattering. This allows us to deal with long-range potential
scattering as short-range potential scattering. An explicit expression of the
scattering wave function for long-range potential scattering is presented, in
which the scattering wave function is represented by the tortoise coordinate
and the scattering phase shift. We show that the long-range potential
scattering wave function is just the short-range potential scattering wave
function with a replacement of a common coordinate by a tortoise coordinate.
The approach applies not only to scattering but also applies to bound states.
Furthermore, in terms of tortoise coordinates, we suggest a classification
scheme for potentials. We also discuss the duality between tortoise coordinates.}


\usepackage{titletoc}   
\titlecontents{section}[0.6em]{\vspace{.25\baselineskip}}
             {\thecontentslabel\quad}{}
             {\hspace{.5em}\titlerule*[10pt]{$\cdot$}\contentspage}
\titlecontents{subsection}[2.1em]{\vspace{.25\baselineskip}}
             {\thecontentslabel\quad}{}
             {\hspace{.5em}\titlerule*[10pt]{$\cdot$}\contentspage}
\titleformat*{\subsection}{\bfseries}
\titleformat*{\subsubsection}{\bfseries}
\numberwithin{subsubsection}{section}
\renewcommand\thesubsubsection{\Alph{subsubsection}}
\begin{document}
\nolinenumbers
\setcounter{tocdepth}{2}
\tableofcontents



\section{Introduction}

Scattering theory is an important issue in both physics and mathematics
\cite{friedrich2015scattering,reed1979scattering,taylor2012scattering}. For
short-range potential scattering, a general theory has been established. For
long-range potential scattering, however, there is no general treatment. The
reason why it is more difficult to establish a general theory for long-range
potential scattering than that for short-range potential scattering is that
the scattering boundary conditions for all short-range potential scatterings
are the same, but for different long-range potentials are different. In this
paper, inspired by general relativity, we suggest a general theory for
long-range potential scattering by introducing tortoise coordinates. The
tortoise coordinate allows us to convert a long-range potential to a
short-range potential. Then a general treatment for long-range potentials can
also be established similar to that of short-range potentials.

\subsubsection{Potential scattering} 
The scattering boundary condition is determined
by the large-distance asymptotic solution of the radial equation
\begin{equation}
\frac{d^{2}u_{l}\left(  r\right)  }{dr^{2}}+\left[  k^{2}-\frac{l\left(
l+1\right)  }{r^{2}}-V\left(  r\right)  \right]  u_{l}\left(  r\right)  =0.
\label{radialeq}%
\end{equation}

For short-range potentials, the large-distance asymptotic behavior is
dominated by the centrifugal potential $\frac{l\left(  l+1\right)  }{r^{2}}$
and the large-distance asymptotics of the radial equation (\ref{radialeq}) is%
\begin{equation}
\frac{d^{2}u_{l}\left(  r\right)  }{dr^{2}}+\left[  k^{2}-\frac{l\left(
l+1\right)  }{r^{2}}\right]  u_{l}\left(  r\right)  \overset{r\rightarrow
\infty}{\sim}0. \label{uk2inf}%
\end{equation}
This asymptotic equation is independent of the potential $V\left(  r\right)
$. Consequently, the scattering boundary conditions for all short-range
potentials are the same and short-range potential scattering can be treated
uniformly \cite{liu2014scattering}.

For long-range potentials, however, the large-distance asymptotic behavior is
dominated by the potential $V\left(  r\right)  $ and the large-distance
asymptotics of the radial equation (\ref{radialeq}) is
\begin{equation}
\frac{d^{2}u_{l}\left(  r\right)  }{dr^{2}}+\left(  k^{2}-V\left(  r\right)
\right)  u_{l}\left(  r\right)  \overset{r\rightarrow\infty}{\sim}0.
\end{equation}
This asymptotic equation depends on the potential $V\left(  r\right)  $.
Consequently, the scattering boundary condition depends on the potential and
in general different potentials lead to different scattering boundary
conditions. This is just the reason why it is difficult to obtain a uniform
scattering boundary condition for long-range potential scattering.

\subsubsection{Tortoise coordinates} Inspect a result in general relativity: in the
Schwarzschild spacetime, the tortoise coordinate can convert the
large-distance asymptotic equation of a long-range potential to that of a
short-range potential.

The Schwarzschild spacetime is described by the metric $ds^{2}=-f\left(
r\right)  dt^{2}+\frac{1}{f\left(  r\right)  }dr^{2}+r^{2}d\Omega^{2}$ with
$f\left(  r\right)  =1-\frac{2M}{r}$. The radial Klein-Gordon equation in the
Schwarzschild spacetime is%
\begin{equation}
\left(  1-\frac{2M}{r}\right)  \frac{d}{dr}\left(  1-\frac{2M}{r}\right)
\frac{du_{l}\left(  r\right)  }{dr}+\left\{  \omega^{2}-\left(  1-\frac{2M}%
{r}\right)  \left[  \frac{l\left(  l+1\right)  }{r^{2}}+\frac{2M}{r^{3}%
}\right]  \right\}  u_{l}\left(  r\right)  =0. \label{KGeq}%
\end{equation}
The large-distance asymptotics of the radial Klein-Gordon equation is%
\begin{equation}
\frac{d^{2}u_{l}\left(  r\right)  }{dr^{2}}+\left(  \omega^{2}+\frac
{4M\omega^{2}}{r}\right)  u_{l}\left(  r\right)  \overset{r\rightarrow
\infty}{\sim}0.
\end{equation}
This is a radial equation with a long-range potential $\frac{4M\omega^{2}}{r}$.

By introducing the tortoise coordinate $r_{\ast}=\int^{r}\frac{1}{f\left(
r\right)  }dr=r+2M\ln\left(  r-2M\right)  $, the radial-time part of the
Schwarzschild spacetime becomes conformally flat, $ds^{2}=f\left(  r\right)
\left(  -dt^{2}+dr_{\ast}^{2}\right)  +r^{2}d\Omega^{2}$, and the
large-distance asymptotics of the radial Klein-Gordon equation (\ref{KGeq})
under the tortoise coordinate becomes \cite{li2018scalar,li2019scattering}
\begin{equation}
\frac{d^{2}u_{l}\left(  r_{\ast}\right)  }{dr_{\ast}^{2}}+\left[  \omega
^{2}-\frac{l\left(  l+1\right)  }{r^{2}}\right]  u_{l}\left(  r_{\ast}\right)
\overset{r\rightarrow\infty}{\sim}0.
\end{equation}
This is just the analogue of the large-distance asymptotic equation of
short-range potential scattering, Eq. (\ref{uk2inf}).

In a word, in the Schwarzschild spacetime the tortoise coordinate converts the
long-range potential scattering to short-range potential scattering. Similar
treatments also applied to the Reissner-Nordstr\"{o}m spacetime
\cite{li2021scalar}.

\subsubsection{Inspiration} In general relativity, as shown above, the tortoise
coordinate converts a long-range potential to a short-range potential. This
inspires us to introduce tortoise coordinates to convert a long-range
potential to a short-range potential in the scattering problem. Once a
long-range potential is converted to a short-range potential, similar to the
treatment for short-range potential scattering, we can also establish a
general theory for long-range potential scattering.

In this paper, we establish a general theory for potential scattering by
introducing the tortoise coordinate which converts\ a long-range potential to
a short-range potential. Under the tortoise coordinate, the large-distance
asymptotic behaviors of all long-range potential scattering are the same.
Recalling that the reason why one can establish a general theory for
short-range potential scattering is that the large-distance asymptotic
behaviors of all short-range potentials are the same, we, in virtue of the
tortoise coordinate, can establish a general treatment for long-range
potential scattering.

Functions can be classified in terms of their asymptotics
\cite{blanton2012introduction}. Wave equations with different long-range
potentials have different asymptotic wave functions, which allows us to
classify long-range potentials by the corresponding asymptotic wave function.
Concretely, in the classification scheme suggested in the present paper, a
long-range potential is converted to a short-range potential by the tortoise
coordinate which is determined by the potential. Different long-range
potentials correspond to different tortoise coordinates, so the classification
of tortoise coordinates classifies the potentials. Especially, a short-range
potential under the tortoise coordinate is just the short-range potential
itself and the asymptotic wave functions for all short-range potentials are
the same, so in this classification scheme, all short-range potentials are
classified into one type.

It is worthy to note here that the result obtained in the present paper not
only applies to scattering but also applies to potentials which possesses only
bound states.

It is revealed that there is a duality in various physical systems
\cite{li2021duality}. We show the duality for tortoise coordinates and for
asymptotic wave functions. We discuss the relation of the classification of
potentials and the duality.

Long-range potential scattering is an important subject in the scattering
theory \cite{Matveev20021648}. There are many studies on long-range potential
scattering, such as the asymptotic completeness of modified wave operators
\cite{derezinski1997long}, the low-energy expansion of the phase shift
\cite{ali1977contribution}, the expansion of the scattering phase shift at
$k=0$ \cite{levy1963low}, the low-energy expansion of the partial-wave Jost
function \cite{willner2006low}, the low-energy scattering theory
\cite{bencze1987low}, the scattering length and the effective range
\cite{ouerdane2004note}, the late-time dynamics of the wave equation
\cite{hod2013scattering}, the Schr\"{o}dinger operator with long-range
electrostatic and magnetic potentials \cite{roux2003scattering}, the spectral
properties of the corresponding long-range potential scattering matrix of the
Schr\"{o}dinger operator \cite{yafaev1998scattering}, the Gell-Mann-Goldberger
formula for long-range potentials \cite{zorbas1976gell}, and the classical
long-range potential scattering \cite{herbst1974classical}. In Ref.
\cite{lavine1970scattering}, the author studies the long-range potential
scattering by proposing a certain weakening of the standard criterion. There
are also studies on scattering including both long-range and short range
potentials, such as the quasi-classical limit of quantum mechanical scattering
for short-range potentials and long-range potentials
\cite{yajima1979quasi,yajima1981quasi}, the short-range and long-range quantum
mechanical scattering \cite{enss1978asymptotic,enss1979asymptotic}, and the
low-energy scattering by a potential consisting of a long-range part and a
short-range part \cite{hinckelmann1971low,ganas1972theory}. For short-range
potential scattering, a rigorous treatment without the large-distance
asymptotics approximation is proposed
\cite{liu2014scattering,li2016scattering}. The duality discuss in the present
paper also exists in various physical systems, such as in the scalar field
\cite{li2021duality} and in the Gross--Pitaevskii equation
\cite{liu2021exactly}.

The tortoise coordinate is first introduced in general relativity and is
widely used in black hole physics \cite{damour1976black}. For the
Schwarzschild spacetime, the Eddington--Finkelstein coordinate is established
on the tortoise coordinate which\ is convenient to describe the ingoing and
outgoing waves \cite{finkelstein1958past,damour1976black}. In Ref.
\cite{hemming2001hawking}, in order to study the Hawking radiation in anti--de
Sitter space, the authors introduce the generalized tortoise coordinate for
the AdS black hole. In Ref. \cite{papantonopoulos2009hawking}, by using the
tortoise coordinate, the radial wave equation in gravitational backgrounds of
a constant negative curvature is simplified.

In section \ref{Tortoisecoordinates}, we\ introduce tortoise coordinates to
convert long-range potentials to short-range potentials. In section
\ref{Classification}, we suggest a classification scheme for potentials in
terms of the tortoise coordinate. In section \ref{phaseshift}, we provide a
uniform expression for long-range potential scattering. In section
\ref{Alternative}, we give an alternative expression of tortoise coordinates.
Some examples are given in sections \ref{Examples}. In section \ref{duality},
we give duality relations for tortoise coordinates and for asymmetric wave
functions. The conclusions are summarized in section \ref{Conclusion}.

\section{Tortoise coordinates: Converting long-range potentials to short-range
potentials \label{Tortoisecoordinates}}

In this section, inspired by general relativity, we introduce tortoise
coordinates and show that a long-range potential\ can be converted to a
short-range potential under the tortoise coordinate.

In the following, potentials are divided into two types to be considered:
potentials vanishing at $r\rightarrow\infty$, which have both scattering
states and bound states and potentials nonvanishing at $r\rightarrow\infty$,
which have only bound states. It can be seen that all kinds of potentials are
converted to short-range potentials including potentials which have only bound states.

\subsection{Potentials vanishing at $r\rightarrow\infty$ \label{mpowers}}

There are two kinds of potentials vanishing at $r\rightarrow\infty$: the
long-range potential, e.g., the Coulomb potential, and the short-range
potential, e.g., the Yukawa potential.

In scattering theory, long-range potential scattering cannot be uniformly
treated, because the large-distance asymptotic behaviors are different for
different long-range potentials. On the contrary, a general theory has been
established for short-range potential scattering, because the large-distance
asymptotic behaviors for all short-range potentials are the same. In the
following, we convert long-range potential scattering to short-range potential
scattering by introducing tortoise coordinates. This allow us to establish a
general theory for long-range potential scattering.

\begin{tcolorbox}[boxrule=0pt,
  boxsep=0pt,
  colback={lightgray1},
  enhanced jigsaw,
  borderline west={3pt}{0pt}{lightgray2},
  sharp corners,
  before skip=10pt,
  after skip=10pt,
breakable,]
\textit{For a potential }$V\left(  r\right)  $\textit{ which vanishes at
}$r\rightarrow\infty$\textit{, by introducing the tortoise coordinate}%
\begin{equation}
r_{\ast}=r-\sum_{\eta=1}^{N}\sigma_{\eta}\int^{r}\left(  \frac{V\left(
r\right)  }{k^{2}}\right)  ^{\eta}dr, \label{torcor1}%
\end{equation}
\textit{where }$\sigma_{\eta}=\Gamma\left(  \eta-1/2\right)  /\left(
2\sqrt{\pi}\eta!\right)  $\textit{, there must exist a non-negative integer
}$N$\textit{ or }$N\rightarrow\infty$\textit{, so that the radial equation
(\ref{radialeq}) with the potential }$V\left(  r\right)  $\textit{ can be
converted to a large-distance asymptotic radial equation of the potential
}$2\sigma_{N+1}\left(  \frac{V\left(  r\right)  }{k^{2}}\right)  ^{N}V\left(
r\right)  $\textit{ which is a short-range potential decreasing faster than
}$1/r$\textit{: }%
\begin{equation}
\frac{d^{2}u_{l}\left(  r_{\ast}\right)  }{dr_{\ast}^{2}}+\left[  k^{2}%
-\frac{l\left(  l+1\right)  }{r^{2}}-2\sigma_{N+1}\left(  \frac{V\left(
r\right)  }{k^{2}}\right)  ^{N}V\left(  r\right)  \right]  u_{l}\left(
r_{\ast}\right)  \sim0. \label{shorteq}%
\end{equation}
\textit{The large-distance asymptotics of Eq. (\ref{shorteq}) under the
tortoise coordinate is}%
\begin{equation}
\frac{d^{2}u_{l}\left(  r_{\ast}\right)  }{dr_{\ast}^{2}}+k^{2}u_{l}\left(
r_{\ast}\right)  \sim0 \label{reqasy1}%
\end{equation}
\textit{and the solution of Eq. (\ref{reqasy1}) is}%
\begin{equation}
u_{l}\left(  r_{\ast}\right)  \sim e^{\pm ikr_{\ast}}. \label{radialwfrstar}%
\end{equation}
\end{tcolorbox}

\textit{Proof}. The\ radial equation (\ref{radialeq}) with the potential
$V\left(  r\right)  $ which vanishes at $r\rightarrow\infty$ under the
tortoise coordinate (\ref{torcor1}) becomes%
\begin{equation}
\frac{d^{2}u_{l}\left(  r_{\ast}\right)  }{dr_{\ast}^{2}}-\frac{\sum_{\eta
=1}^{N}\eta\sigma_{\eta}\left(  \frac{V\left(  r\right)  }{k^{2}}\right)
^{\eta}\frac{V^{\prime}\left(  r\right)  }{V\left(  r\right)  }}{\left[
1-\sum_{\eta=1}^{N}\sigma_{\eta}\left(  \frac{V\left(  r\right)  }{k^{2}%
}\right)  ^{\eta}\right]  ^{2}}\frac{du_{l}\left(  r_{\ast}\right)  }%
{dr_{\ast}}+\frac{1}{\left[  1-\sum_{\eta=1}^{N}\sigma_{\eta}\left(
\frac{V\left(  r\right)  }{k^{2}}\right)  ^{\eta}\right]  ^{2}}\left[
k^{2}-\frac{l\left(  l+1\right)  }{r^{2}}-V\left(  r\right)  \right]
u_{l}\left(  r_{\ast}\right)  =0,
\end{equation}
where%
\begin{equation}
dr_{\ast}=\left[  1-\sum_{\eta=1}^{N}\sigma_{\eta}\left(  \frac{V\left(
r\right)  }{k^{2}}\right)  ^{\eta}\right]  dr \label{rstrar3}%
\end{equation}
is used.

At $r\rightarrow\infty$, the coefficient of $\frac{du_{l}\left(  r_{\ast
}\right)  }{dr_{\ast}}$ vanishes, i.e.
\begin{equation}
\frac{\sum_{\eta=1}^{N}\eta\sigma_{\eta}\left(  \frac{V\left(  r\right)
}{k^{2}}\right)  ^{\eta}\frac{V^{\prime}\left(  r\right)  }{V\left(  r\right)
}}{\left[  1-\sum_{\eta=1}^{N}\sigma_{\eta}\left(  \frac{V\left(  r\right)
}{k^{2}}\right)  ^{\eta}\right]  ^{2}}\sim\frac{1}{2}\frac{V^{\prime}\left(
r\right)  }{k^{2}}\overset{r\rightarrow\infty}{\sim}0.
\end{equation}
Note that $V^{\prime}\left(  r\right)  $ falls off faster than $\frac{1}{r}$
and vanishes at $r\rightarrow\infty$.

At $r\rightarrow\infty$, the coefficient of $u_{l}\left(  r_{\ast}\right)  $
becomes
\begin{equation}
\frac{1}{\left[  1-\sum_{\eta=1}^{N}\sigma_{\eta}\left(  \frac{V\left(
r\right)  }{k^{2}}\right)  ^{\eta}\right]  ^{2}}\left[  k^{2}-\frac{l\left(
l+1\right)  }{r^{2}}-V\left(  r\right)  \right]  \overset{r\rightarrow
\infty}{\sim}k^{2}-\frac{l\left(  l+1\right)  }{r^{2}}-2\sigma_{N+1}\left(
\frac{V\left(  r\right)  }{k^{2}}\right)  ^{N}V\left(  r\right)  +\cdots.
\end{equation}
Clearly, there must exist a value of $N$ so that $2\sigma_{N+1}\left(
\frac{V\left(  r\right)  }{k^{2}}\right)  ^{N}V\left(  r\right)  $ falls
faster than $1/r$. This proves Eq. (\ref{shorteq}).

\subsection{Potentials nonvanishing at $r\rightarrow\infty$ \label{ppowers}}

Potentials nonvanishing at $r\rightarrow\infty$ are long-range distance
potentials which has only bound states. In this section, we show that even
potentials which possess only bound states can also be converted to
short-range potentials by introducing the tortoise coordinate.

\begin{tcolorbox}[boxrule=0pt,
  boxsep=0pt,
  colback={lightgray1},
  enhanced jigsaw,
  borderline west={3pt}{0pt}{lightgray2},
  sharp corners,
  before skip=10pt,
  after skip=10pt,
breakable,]
\textit{For the potential }$V\left(  r\right)  $\textit{ which does not vanish
at }$r\rightarrow\infty$\textit{, by introducing the tortoise coordinate}%
\begin{equation}
r_{\ast}=\frac{1}{4k}\ln\frac{V\left(  r\right)  }{k^{2}}-\sum_{\eta=0}%
^{N}\sigma_{\eta}\int^{r}\left(  \frac{k^{2}}{V\left(  r\right)  }\right)
^{\eta-1/2}dr, \label{torcor2}%
\end{equation}
\textit{there must exist a non-negative integer }$N$\textit{ or }%
$N\rightarrow\infty$\textit{, so that the radial equation (\ref{radialeq})
with the potential }$V\left(  r\right)  $\textit{ can be converted to a
large-distance asymptotic radial equation of the potential }$\frac{1}{2}%
\frac{k^{2}}{V\left(  r\right)  }\frac{V^{\prime}\left(  r\right)  }{\sqrt{V}%
}$\textit{ which is a short-range potential decreasing faster than }%
$1/r$\textit{:}%
\begin{equation}
\frac{d^{2}u_{l}\left(  r_{\ast}\right)  }{dr_{\ast}^{2}}+\left(  -k^{2}%
+\frac{1}{2}\frac{k^{2}}{V\left(  r\right)  }\frac{V^{\prime}\left(  r\right)
}{\sqrt{V\left(  r\right)  }}\right)  u_{l}\left(  r_{\ast}\right)  \sim0.
\label{shorteq2}%
\end{equation}
\textit{The large-distance asymptotics of Eq. (\ref{shorteq2}) under the
tortoise coordinate is}%
\begin{equation}
\frac{d^{2}u_{l}\left(  r_{\ast}\right)  }{dr_{\ast}^{2}}-k^{2}u_{l}\left(
r_{\ast}\right)  \sim0, \label{reqasy2}%
\end{equation}
\textit{and the solution of Eq. (\ref{reqasy2}) is}%
\begin{equation}
u_{l}\left(  r_{\ast}\right)  \sim e^{-kr_{\ast}}. \label{uinf}%
\end{equation}
\end{tcolorbox}

\textit{Proof.} The\ radial equation (\ref{radialeq}) with the potential which
does not vanish at $r\rightarrow\infty$ under the tortoise coordinate
(\ref{torcor2}) becomes%
\begin{align}
\frac{d^{2}u_{l}\left(  r_{\ast}\right)  }{dr_{\ast}^{2}}  &  +\frac
{\sum_{\eta=0}^{N}\left(  \eta-\frac{1}{2}\right)  \sigma_{\eta}\left(
\frac{k^{2}}{V\left(  r\right)  }\right)  ^{\eta-1/2}\frac{V^{\prime}\left(
r\right)  }{V\left(  r\right)  }+\frac{1}{4k}\left[  \frac{V^{\prime\prime
}\left(  r\right)  }{V\left(  r\right)  }-\left(  \frac{V^{\prime}\left(
r\right)  }{V\left(  r\right)  }\right)  ^{2}\right]  }{\left[  -\sum_{\eta
=0}^{N}\sigma_{\eta}\left(  \frac{k^{2}}{V\left(  r\right)  }\right)
^{\eta-1/2}+\frac{1}{4k}\frac{V^{\prime}\left(  r\right)  }{V\left(  r\right)
}\right]  ^{2}}\frac{du_{l}\left(  r_{\ast}\right)  }{dr_{\ast}}\nonumber\\
&  +\frac{1}{\left[  -\sum_{\eta=0}^{N}\sigma_{\eta}\left(  \frac{k^{2}%
}{V\left(  r\right)  }\right)  ^{\eta-1/2}+\frac{1}{4k}\frac{V^{\prime}\left(
r\right)  }{V\left(  r\right)  }\right]  ^{2}}\left[  k^{2}-\frac{l\left(
l+1\right)  }{r^{2}}-V\left(  r\right)  \right]  u_{l}\left(  r\right)  =0,
\end{align}
where
\begin{equation}
dr_{\ast}=\left[  -\sum_{\eta=0}^{N}\sigma_{\eta}\left(  \frac{k^{2}}{V\left(
r\right)  }\right)  ^{\eta-1/2}+\frac{1}{4k}\frac{V^{\prime}\left(  r\right)
}{V\left(  r\right)  }\right]  dr
\end{equation}
is used.

At $r\rightarrow\infty$, the coefficient of $\frac{du_{l}\left(  r_{\ast
}\right)  }{dr_{\ast}}$ vanishes, i.e.
\begin{equation}
\frac{\sum_{\eta=0}^{N}\left(  \eta-\frac{1}{2}\right)  \sigma_{\eta}\left(
\frac{k^{2}}{V\left(  r\right)  }\right)  ^{\eta-1/2}\frac{V^{\prime}\left(
r\right)  }{V\left(  r\right)  }+\frac{1}{4k}\left[  \frac{V^{\prime\prime
}\left(  r\right)  }{V\left(  r\right)  }-\left(  \frac{V^{\prime}\left(
r\right)  }{V\left(  r\right)  }\right)  ^{2}\right]  }{\left[  -\sum_{\eta
=0}^{N}\sigma_{\eta}\left(  \frac{k^{2}}{V\left(  r\right)  }\right)
^{\eta-1/2}+\frac{1}{4k}\frac{V^{\prime}\left(  r\right)  }{V\left(  r\right)
}\right]  ^{2}}\overset{r\rightarrow\infty}{\sim}\frac{1}{2}\sqrt{\frac{k^{2}%
}{V\left(  r\right)  }}\frac{V^{\prime}\left(  r\right)  }{V\left(  r\right)
}\sim0.
\end{equation}
Note that $\frac{1}{2}\sqrt{\frac{k^{2}}{V\left(  r\right)  }}\frac{V^{\prime
}\left(  r\right)  }{V\left(  r\right)  }$ falls off faster than $1/r$ and
then vanishes at $r\rightarrow\infty$.

At $r\rightarrow\infty$, the coefficients of $u_{l}\left(  r_{\ast}\right)  $
becomes
\end{paracol}
\begin{align}
  \qquad \qquad\qquad\quad\frac{1}{\left(  -\sum_{\eta=0}^{N}\sigma_{\eta}\left(  \frac{k^{2}%
}{V\left(  r\right)  }\right)  ^{\eta-1/2}+\frac{1}{4k}\frac{V^{\prime}\left(
r\right)  }{V\left(  r\right)  }\right)  ^{2}}\left[  k^{2}-\frac{l\left(
l+1\right)  }{r^{2}}-V\left(  r\right)  \right]
\overset{r\rightarrow\infty}{\sim}    -k^{2}+\frac{1}{2}\frac{k^{2}}{V\left(
r\right)  }\frac{V^{\prime}\left(  r\right)  }{\sqrt{V\left(  r\right)  }%
}+2\sigma_{N+1}\left(  \frac{k^{2}}{V\left(  r\right)  }\right)
^{N+2}V\left(  r\right)  .
\end{align}
\begin{paracol}{2}
\switchcolumn \noindent Clearly, there must exist a value of $N$ so that $2\sigma_{N+1}\left(
\frac{k^{2}}{V\left(  r\right)  }\right)  ^{N+2}V\left(  r\right)  $ falls
faster than $\frac{1}{2}\frac{k^{2}}{V\left(  r\right)  }\frac{V^{\prime
}\left(  r\right)  }{\sqrt{V\left(  r\right)  }}$ and $\frac{1}{2}\frac{k^{2}%
}{V\left(  r\right)  }\frac{V^{\prime}\left(  r\right)  }{\sqrt{V\left(
r\right)  }}$ falls faster than $1/r$.

It can be seen that the eigenvalue $-k^{2}$ in Eq. (\ref{shorteq2}) is less
than zero, since the potential considered here has only bound states.

\section{Classifying potentials in terms of tortoise coordinates
\label{Classification}}

In this section, we suggest a classification scheme for potentials in terms of
tortoise coordinates. Essentially, this scheme classifies potentials by the
asymptotic behaviors of the corresponding wave functions. The result given by
Eqs. (\ref{radialwfrstar}) and (\ref{uinf}) shows that under the tortoise
coordinate the asymptotic wave functions are the same. The difference between
asymptotic wave functions is reflected in tortoise coordinates which depend on
potentials. Therefore, the classification of tortoise coordinates classifies potentials.

\subsection{Potentials vanishing at $r\rightarrow\infty$
\label{Classificationvanishing}}

In the following, we suggest a classification scheme for potentials in terms
of the tortoise coordinate according to various values of $N$ in the
definition of the tortoise coordinate (\ref{torcor1}).

\begin{tcolorbox}[boxrule=0pt,
  boxsep=0pt,
  colback={lightgray1},
  enhanced jigsaw,
  borderline west={3pt}{0pt}{lightgray2},
  sharp corners,
  before skip=10pt,
  after skip=10pt,
breakable,]
\textit{The potentials which vanish at }$r\rightarrow\infty$\textit{ can be
classified in three types in terms of the value of }$N$\textit{ in the
tortoise coordinate (\ref{torcor1}):}

\begin{enumerate}
\item $N=0$\textit{. The potential corresponding to }$N=0$\textit{ is a
short-range potential satisfying}%
\begin{equation}
\int_{a}^{\infty}\left\vert V\left(  r\right)  \right\vert dr<\infty
,\text{\ }a>0, \label{Vcd1}%
\end{equation}
\textit{i.e., the potential }$V\left(  r\right)  $\textit{ decreases faster
than }$1/r$\textit{ at }$r\rightarrow\infty$\textit{. In this case, the
tortoise coordinate is just the radial coordinate itself: }$r_{\ast}%
=r$\textit{. \ }

\item $N$\textit{ is a positive integer. The potential corresponding to a
positive integer }$N$\textit{ is a long-range potential satisfying}%
\begin{align}
\int_{a}^{\infty}\left\vert V\left(  r\right)  \right\vert r^{A-1}dr  &
<\infty,\text{ \ }a>0,\text{ \ }A=\frac{1}{N+1},\label{cd1}\\
\int_{b}^{\infty}\frac{1}{\left\vert V\left(  r\right)  \right\vert
r^{B+1+\epsilon}}dr  &  <\infty,\text{ \ }b>0,\text{ \ }B=\frac{1}{N},\text{
}\epsilon\sim0^{+}, \label{cd2}%
\end{align}
\textit{i.e., the potential }$V\left(  r\right)  $\textit{ decreases faster
than }$1/r^{1/\left(  N+1\right)  }$\textit{ and slower than or equally to
}$1/r^{1/N}$\textit{\ at }$r\rightarrow\infty$\textit{. Different values of
}$N$\textit{ correspond to different long-range potentials with different
potential ranges.}

\item $N\rightarrow\infty$\textit{. The potential corresponding to
}$N\rightarrow\infty$\textit{ is another type of long-range potentials
satisfying}%
\begin{align}
\int_{a}^{\infty}\frac{\left\vert V\left(  r\right)  \right\vert }{r}dr  &
<\infty,\text{ \ }a>0,\label{Vcd12}\\
\int_{b}^{\infty}\frac{1}{\left\vert V\left(  r\right)  \right\vert
r^{1+\epsilon}}dr  &  <\infty,\text{ \ }b>0,\text{ }\epsilon\sim0^{+}.
\label{Vcd2}%
\end{align}
\textit{The tortoise coordinate (\ref{torcor1}) in this case becomes}%
\begin{equation}
r_{\ast}=\int^{r}\sqrt{1-\frac{V\left(  r\right)  }{k^{2}}}dr. \label{torcor4}%
\end{equation}
\textit{Under the tortoise coordinate (\ref{torcor4}), the radial equation
(\ref{radialeq}) becomes}%
\begin{equation}
\frac{d^{2}u_{l}\left(  r_{\ast}\right)  }{dr_{\ast}^{2}}+\left[  k^{2}%
-\frac{l\left(  l+1\right)  }{r^{2}}\frac{1}{1-V\left(  r\right)  /k^{2}%
}\right]  u_{l}\left(  r_{\ast}\right)  =0. \label{shorteq4}%
\end{equation}

\end{enumerate}
\end{tcolorbox}

A special case of the potential corresponding to $N=0$ is the negative power
potential $1/r^{\alpha}$ with $\alpha>1$; a special case of the potential
corresponding to $N$ equalling a finite positive integer is the negative power
potential $1/r^{\alpha}$ with $0<\alpha\leq1$; two special cases of the
potentials corresponding to $N\rightarrow\infty$ are the constant potential
and the potential $V\left(  r\right)  \sim1/\ln r$.

In the following, we prove the above statements.

\textit{Proof.} First, we write the radial wave function as%
\begin{equation}
u_{l}\left(  r\right)  =e^{\pm ikr_{\ast}}e^{h\left(  r\right)  }. \label{ulr}%
\end{equation}
Substituting Eq. (\ref{ulr}) into the radial equation (\ref{radialeq}) gives
an equation of $h\left(  r\right)  $,%
\begin{equation}
h^{\prime\prime}\left(  r\right)  +h^{\prime}\left(  r\right)  ^{2}%
+2\frac{\left(  e^{\pm ikr_{\ast}}\right)  ^{\prime}}{e^{\pm ikr_{\ast}}%
}h^{\prime}\left(  r\right)  =-k^{2}+\frac{l\left(  l+1\right)  }{r^{2}%
}+V\left(  r\right)  -\frac{\left(  e^{\pm ikr_{\ast}}\right)  ^{\prime\prime
}}{e^{\pm ikr_{\ast}}}. \label{heq}%
\end{equation}
The asymptotics of Eq. (\ref{heq}) at $r\rightarrow\infty$ is%
\begin{equation}
\pm2ikh^{\prime}\left(  r\right)  \sim2\sigma_{N+1}\left(  \frac{V\left(
r\right)  }{k^{2}}\right)  ^{N}V\left(  r\right)  .
\end{equation}
Solving this asymptotic equation gives%
\begin{equation}
h\left(  r\right)  \sim\pm\sigma_{N+1}\frac{k}{i}\int^{r}\left(
\frac{V\left(  r\right)  }{k^{2}}\right)  ^{N+1}dr. \label{hr}%
\end{equation}
To satisfy\ Eq. (\ref{radialwfrstar}), we require that%
\begin{equation}
h\left(  r\right)  \overset{r\rightarrow\infty}{\sim}0. \label{hrinf}%
\end{equation}
Eqs. (\ref{hr}) and (\ref{hrinf}) require that $V\left(  r\right)  $ decreases
faster than $1/r^{\frac{1}{N+1}}$ at $r\rightarrow\infty$, so%
\begin{align}
\int_{a}^{\infty}\left\vert V\left(  r\right)  \right\vert r^{\frac{1}{N+1}%
-1}dr  &  =\int_{a}^{c}\left\vert V\left(  r\right)  \right\vert r^{\frac
{1}{N+1}-1}dr+\int_{c}^{\infty}\left\vert V\left(  r\right)  \right\vert
r^{\frac{1}{N+1}-1}dr\nonumber\\
&  \leq\int_{a}^{c}\left\vert V\left(  r\right)  \right\vert r^{\frac{1}%
{N+1}-1}dr+\int_{c}^{\infty}\frac{1}{r^{\frac{1}{N+1}+\delta}}r^{\frac{1}%
{N+1}-1}dr\nonumber\\
&  =\int_{a}^{c}\left\vert V\left(  r\right)  \right\vert r^{\frac{1}{N+1}%
-1}dr+\int_{c}^{\infty}\frac{1}{r^{1+\delta}}dr, \label{v1}%
\end{align}
where $\delta>0$. Then we have $\int_{a}^{\infty}\left\vert V\left(  r\right)
\right\vert r^{A-1}dr<\infty$ with\ $a>0$ and\ $A=\frac{1}{N+1}$, since
Eq.(\ref{v1}) is finite.

$N=0$ gives the condition (\ref{Vcd1}), $N$ equaling a nonzero finite integer
gives the condition (\ref{cd2}), and $N\rightarrow\infty$ gives the condition
(\ref{Vcd2}), respectively.

Second, we write the radial wave function as%
\begin{equation}
u_{l}\left(  r\right)  =\exp\left(  \pm ik\left[  r_{\ast}+\sigma_{N}\int%
^{r}\left(  \frac{V\left(  r\right)  }{k^{2}}\right)  ^{N}dr\right]  \right)
e^{g\left(  r\right)  }. \label{ulr2}%
\end{equation}
Substituting Eq. (\ref{ulr2}) into the radial equation (\ref{radialeq}) gives
an equation of $g\left(  r\right)  $,%
\end{paracol}
\begin{align}
g^{\prime\prime}\left(  r\right)  +g^{\prime}\left(  r\right)  ^{2}%
+2\frac{\left\{  \exp\left(  \pm ik\left[  r_{\ast}+\sigma_{N}\int^{r}\left(
\frac{V\left(  r\right)  }{k^{2}}\right)  ^{N}dr\right]  \right)  \right\}
^{\prime}}{\exp\left(  \pm ik\left[  r_{\ast}+\sigma_{N}\int^{r}\left(
\frac{V\left(  r\right)  }{k^{2}}\right)  ^{N}dr\right]  \right)  }g^{\prime
}\left(  r\right) 
=   -k^{2}+\frac{l\left(  l+1\right)  }{r^{2}}+V\left(  r\right)
-\frac{\left\{  \exp\left(  \pm ik\left[  r_{\ast}+\sigma_{N}\int^{r}\left(
\frac{V\left(  r\right)  }{k^{2}}\right)  ^{N}dr\right]  \right)  \right\}
^{\prime\prime}}{\exp\left(  \pm ik\left[  r_{\ast}+\sigma_{N}\int^{r}\left(
\frac{V\left(  r\right)  }{k^{2}}\right)  ^{N}dr\right]  \right)  }.
\label{weq}%
\end{align}
\begin{paracol}{2}
\switchcolumn
\noindent The asymptotics of Eq. (\ref{weq}) at $r\rightarrow\infty$ reads%
\begin{equation}
\pm2ikg^{\prime}\left(  r\right)  \sim2\sigma_{N}\left(  \frac{V\left(
r\right)  }{k^{2}}\right)  ^{N-1}V\left(  r\right)  .
\end{equation}
Solving this asymptotic equation gives%
\begin{equation}
g\left(  r\right)  \sim\pm\frac{\sigma_{N}}{ik}\int^{r}\left(  \frac{V\left(
r\right)  }{k^{2}}\right)  ^{N}dr. \label{gr}%
\end{equation}
To satisfy Eq. (\ref{radialwfrstar}), at $r\rightarrow\infty$, the factor
$g\left(  r\right)  $ must contribute, i.e.,
\begin{equation}
g\left(  r\right)  \overset{r\rightarrow\infty}{\nsim}0. \label{grnot0}%
\end{equation}
Eqs. (\ref{gr}) and (\ref{grnot0}) require that $V\left(  r\right)  $ must
decrease slower than or equally to $1/r^{1/N}$ at\ $r\rightarrow\infty$, so%
\begin{align}
\int_{b}^{\infty}\frac{1}{\left\vert V\left(  r\right)  \right\vert
r^{\frac{1}{N}+1+\epsilon}}dr  &  =\int_{b}^{c}\frac{1}{\left\vert V\left(
r\right)  \right\vert r^{\frac{1}{N}+1+\epsilon}}dr+\int_{c}^{\infty}\frac
{1}{\left\vert V\left(  r\right)  \right\vert r^{\frac{1}{N}+1+\epsilon}%
}dr\nonumber\\
&  \leq\int_{b}^{c}\frac{1}{\left\vert V\left(  r\right)  \right\vert
r^{\frac{1}{N}+1+\epsilon}}dr+\int_{c}^{\infty}\frac{1}{\frac{1}{r^{\frac
{1}{N}-\delta}}r^{\frac{1}{N}+1+\epsilon}}dr\nonumber\\
&  =\int_{b}^{c}\frac{1}{\left\vert V\left(  r\right)  \right\vert r^{\frac
{1}{N}+1+\epsilon}}dr+\int_{c}^{\infty}\frac{1}{r^{1+\epsilon+\delta}}dr,
\label{v2}%
\end{align}
where $\delta\geq0$. Then we have $\int_{b}^{\infty}\frac{1}{\left\vert
V\left(  r\right)  \right\vert r^{B+1+\epsilon}}dr<\infty$ with $b>0$%
,\ $B=1/N$, and $\epsilon\sim0^{+}$, since Eq.(\ref{v1}) is finite.

$N$ equaling a nonzero finite integer gives the condition (\ref{cd1}) and
$N\rightarrow\infty$ gives the condition (\ref{Vcd12}), respectively.
Especially, for $N\rightarrow\infty$, substituting the tortoise coordinate
(\ref{torcor4}) into Eq. (\ref{radialeq}) gives%
\begin{equation}
\frac{d^{2}u_{l}\left(  r_{\ast}\right)  }{dr_{\ast}^{2}}-\frac{V^{\prime
}\left(  r\right)  }{2k^{2}\left(  1-\frac{V\left(  r\right)  }{k^{2}}\right)
^{3/2}}\frac{du_{l}\left(  r_{\ast}\right)  }{dr_{\ast}}+\left[  k^{2}%
-\frac{l\left(  l+1\right)  }{r^{2}}\frac{1}{1-\frac{V\left(  r\right)
}{k^{2}}}\right]  u_{l}\left(  r_{\ast}\right)  =0.
\end{equation}
At $r\rightarrow\infty$, the coefficient of $du_{l}\left(  r_{\ast}\right)
/dr_{\ast}$ falls off faster than $1/r$ and then vanishes; at the meantime,
$\frac{\left(  l+1\right)  l}{r^{2}\left(  1-V\left(  r\right)  /k^{2}\right)
}$ is a a short-range potential. This proves Eq. (\ref{shorteq4}).

{\bfseries{Remark}}. In this classification scheme, the potentials decaying faster
than $\frac{1}{r}$ are classified into one class, for their asymptotic
scattering wave function are the same.

In literature, there are two classification schemes: (1) in terms of the
comparison of the potential and the centrifugal potential $\frac{1}{r^{2}}$,
and (2) in terms of the asymptotic behavior of scattering wave functions. In
the scheme (1),\ the short-range potential decays faster than $\frac{1}{r^{2}%
}$, while in the scheme (2), the short-range potential decays faster than
$\frac{1}{r}$. The scheme (1) corresponds to the requirement
\begin{equation}
\int_{a}^{\infty}\left\vert V\left(  r\right)  \right\vert rdr<\infty,\text{
\ }a>0. \label{I.1.3}%
\end{equation}
The scheme (2) corresponds to the requirement%
\begin{equation}
\int_{a}^{\infty}\left\vert V\left(  r\right)  \right\vert dr<\infty,\text{
\ }a>0. \label{I.1.1}%
\end{equation}
Usually we adopt the scheme (1) for avoiding unnecessary complications
\cite{chadan2012inverse}. However, the essential classification scheme is the
scheme (2) \cite{chadan2012inverse,burke2011r}.

\subsection{Potentials nonvanishing at $r\rightarrow\infty$
\label{Classificationnonvanishing}}

In the following, as above, we classify the potential by the tortoise
coordinate according to various values of $N$ in the definition of the
tortoise coordinate (\ref{torcor2}).

\begin{tcolorbox}[boxrule=0pt,
  boxsep=0pt,
  colback={lightgray1},
  enhanced jigsaw,
  borderline west={3pt}{0pt}{lightgray2},
  sharp corners,
  before skip=10pt,
  after skip=10pt,
breakable,]
\textit{The potentials which does not vanish at }$r\rightarrow\infty$\textit{
can be classified in three types in terms of the value of }$N$\textit{ in the
tortoise coordinate (\ref{torcor2}):}

\begin{enumerate}
\item $N=0$\textit{. The potential corresponding to }$N=0$\textit{ satisfies}%
\begin{equation}
\int_{b}^{\infty}\frac{1}{\left\vert V\left(  r\right)  \right\vert
}rdr<\infty, \label{Vpcd1}%
\end{equation}
\textit{i.e., the potential }$V\left(  r\right)  $ \textit{increases faster
than }$r^{2}$\textit{ at }$r\rightarrow\infty$\textit{. We might call it the
superlong-range potential. The tortoise coordinate (\ref{torcor1}) in this
case becomes}%
\begin{equation}
r_{\ast}=\frac{1}{4k}\ln\frac{V\left(  r\right)  }{k^{2}}+\int^{r}\sqrt
{\frac{V\left(  r\right)  }{k^{2}}}dr.
\end{equation}

\item $N$\textit{ is a positive integer. The potential corresponding to a
positive integer }$N$\textit{ is a long-range potentials satisfying}%
\begin{align}
\int_{b}^{\infty}\frac{1}{\left\vert V\left(  r\right)  \right\vert }%
r^{A-1}dr  &  <\infty,\text{ \ }b>0,\text{ \ }A=\frac{1}{N+1/2},\label{cd3}\\
\int_{a}^{\infty}\left\vert V\left(  r\right)  \right\vert \frac
{1}{r^{B+1+\epsilon}}dr  &  <\infty,\text{ \ }a>0,\text{ \ }B=\frac{1}%
{N-1/2},\text{ }\epsilon\sim0^{+}, \label{cd4}%
\end{align}
\textit{i.e., the potential }$V\left(  r\right)  $\textit{ increases faster
than }$r^{1/\left(  N+1/2\right)  }$\textit{ and slower than or equally to
}$r^{1/\left(  N-1/2\right)  }$\textit{\ at }$r\rightarrow\infty$\textit{.
Different values of }$N$\textit{ correspond to different long-range potentials
with different potential ranges.}

\item $N\rightarrow\infty$\textit{. The potential corresponding to
}$N\rightarrow\infty$\textit{ satisfies}%
\begin{align}
\int_{b}^{\infty}\frac{1}{\left\vert V\left(  r\right)  \right\vert }r^{-1}dr
&  <\infty,\text{ \ }b>0,\label{Vpcd2}\\
\int_{a}^{\infty}\left\vert V\left(  r\right)  \right\vert \frac
{1}{r^{1+\epsilon}}dr  &  <\infty,\text{ \ }a>0,\text{ }\epsilon\sim0^{+}.
\label{Vpcd3}%
\end{align}
\textit{The tortoise coordinate (\ref{torcor2}) in this case becomes}%
\begin{equation}
r_{\ast}=\frac{1}{4k}\ln\frac{V\left(  r\right)  }{k^{2}}+\int^{r}%
dr\sqrt{\frac{V\left(  r\right)  }{k^{2}}}\sqrt{1-\frac{k^{2}}{V\left(
r\right)  }}. \label{torcor5}%
\end{equation}
\textit{Under the tortoise coordinate (\ref{torcor5}), the radial equation
(\ref{radialeq}) becomes}%
\begin{equation}
\frac{d^{2}u_{l}\left(  r_{\ast}\right)  }{dr_{\ast}^{2}}+\left[  -k^{2}%
-\frac{l\left(  l+1\right)  }{\left(  \frac{V\left(  r\right)  }{k^{2}%
}-1\right)  r^{2}}+\frac{V^{\prime}\left(  r\right)  /k}{2\frac{V\left(
r\right)  }{k^{2}}\sqrt{\frac{V\left(  r\right)  }{k^{2}}-1}}\right]
u_{l}\left(  r_{\ast}\right)  =0. \label{shorteq5}%
\end{equation}

\end{enumerate}
\end{tcolorbox}

A special case of the potential corresponding to $N=0$ is $r^{\alpha}$ with
$\alpha>2$; a special case of the potential corresponding to a finite positive
integer $N$ is the positive power potential $r^{\alpha}$ with $0<\alpha\leq2$;
two special cases of the potential corresponding to $N\rightarrow\infty$ are
constant potential and the potential $V\left(  r\right)  \sim\ln r$.

In the following, we prove the above statement.

\textit{Proof.} First, we write the radial wave function as%
\begin{equation}
u_{l}\left(  r\right)  =e^{-kr_{\ast}}e^{q\left(  r\right)  }. \label{ulrq}%
\end{equation}
Substituting Eq. (\ref{ulrq}) into the radial equation (\ref{radialeq}) gives
an equation of $q\left(  r\right)  $,%
\begin{equation}
q^{\prime\prime}\left(  r\right)  +q^{\prime}\left(  r\right)  ^{2}%
+2\frac{\left(  e^{-kr_{\ast}}\right)  ^{\prime}}{e^{-kr_{\ast}}}q^{\prime
}\left(  r\right)  =-k^{2}+\frac{l\left(  l+1\right)  }{r^{2}}+V\left(
r\right)  -\frac{\left(  e^{-kr_{\ast}}\right)  ^{\prime\prime}}{e^{-kr_{\ast
}}}. \label{qeq}%
\end{equation}
The asymptotics of Eq. (\ref{qeq}) at $r\rightarrow\infty$ is%
\begin{equation}
2\sqrt{V\left(  r\right)  }q^{\prime}\left(  r\right)  \sim-2\sigma
_{N+1}\left(  \frac{k^{2}}{V\left(  r\right)  }\right)  ^{N+1}V\left(
r\right)  .
\end{equation}
Solving this asymptotic equation gives%
\begin{equation}
q\left(  r\right)  \sim-k\sigma_{N+1}\int^{r}\left(  \frac{k^{2}}{V\left(
r\right)  }\right)  ^{N+1/2}dr. \label{qr}%
\end{equation}
To satisfy\ Eq. (\ref{uinf}), we require that
\begin{equation}
q\left(  r\right)  \overset{r\rightarrow\infty}{\sim}0. \label{qr0}%
\end{equation}
Eqs. (\ref{qr}) and (\ref{qr0}) require that $V\left(  r\right)  $ increases
faster than $r^{\frac{1}{N+1/2}}$, so%
\begin{align}
\int_{b}^{\infty}\frac{1}{\left\vert V\left(  r\right)  \right\vert }%
r^{\frac{1}{N+1/2}-1}dr  &  =\int_{b}^{c}\frac{1}{\left\vert V\left(
r\right)  \right\vert }r^{\frac{1}{N+1/2}-1}dr+\int_{c}^{\infty}\frac
{1}{\left\vert V\left(  r\right)  \right\vert }r^{\frac{1}{N+1/2}%
-1}dr\nonumber\\
&  \leq\int_{b}^{c}\frac{1}{\left\vert V\left(  r\right)  \right\vert
}r^{\frac{1}{N+1/2}-1}dr+\int_{c}^{\infty}\frac{1}{r^{\frac{1}{N+1/2}+\delta}%
}r^{\frac{1}{N+1/2}-1}dr\nonumber\\
&  =\int_{b}^{c}\frac{1}{\left\vert V\left(  r\right)  \right\vert }%
r^{\frac{1}{N+1/2}-1}dr+\int_{c}^{\infty}\frac{1}{r^{1+\delta}}dr, \label{v3}%
\end{align}
where $\delta>0$. Then we have $\int_{b}^{\infty}\frac{1}{\left\vert V\left(
r\right)  \right\vert }r^{A-1}dr<\infty$ with\ $b>0$ and\ $A=\frac{1}{N+1/2}$,
since Eq.(\ref{v3}) is finite.

$N=0$ gives the condition (\ref{Vpcd1}); $N$ equaling a nonzero finite
positive integer gives the condition (\ref{cd3}); $N\rightarrow\infty$ gives
the condition (\ref{Vpcd2}), respectively.

Second, we write the radial wave function as%
\begin{equation}
u_{l}\left(  r\right)  =\exp\left(  -kr_{\ast}+k\sigma_{N}\int^{r}\left(
\frac{k^{2}}{V\left(  r\right)  }\right)  ^{N-1/2}dr\right)  e^{p\left(
r\right)  }. \label{ulrq2}%
\end{equation}
Substituting Eq. (\ref{ulrq2}) into the radial equation Eq. (\ref{radialeq})
gives an equation of $p\left(  r\right)  $,%
\end{paracol}
\begin{align}
p^{\prime\prime}\left(  r\right)  +p^{\prime}\left(  r\right)  ^{2}%
+2\frac{\left[  \exp\left(  -kr_{\ast}+k\sigma_{N}\int^{r}\left(  \frac{k^{2}%
}{V\left(  r\right)  }\right)  ^{N-1/2}dr\right)  \right]  ^{\prime}}%
{\exp\left(  -kr_{\ast}+k\sigma_{N}\int^{r}\left(  \frac{k^{2}}{V\left(
r\right)  }\right)  ^{N-1/2}dr\right)  }p^{\prime}\left(  r\right) 
=   -k^{2}+\frac{l\left(  l+1\right)  }{r^{2}}+V\left(  r\right)
-\frac{\left[  \exp\left(  -kr_{\ast}+k\sigma_{N}\int^{r}\left(  \frac{k^{2}%
}{V\left(  r\right)  }\right)  ^{N-1/2}dr\right)  \right]  ^{\prime\prime}%
}{\exp\left(  -kr_{\ast}+k\sigma_{N}\int^{r}\left(  \frac{k^{2}}{V\left(
r\right)  }\right)  ^{N-1/2}dr\right)  }. \label{peq}%
\end{align}
\begin{paracol}{2}
\switchcolumn \noindent The asymptotics of Eq. (\ref{peq}) at $r\rightarrow\infty$ is%
\begin{equation}
2\sqrt{V\left(  r\right)  }p^{\prime}\left(  r\right)  \sim-2\sigma_{N}\left(
\frac{k^{2}}{V\left(  r\right)  }\right)  ^{N}V\left(  r\right)  .
\end{equation}
Solving this asymptotic equation gives%
\begin{equation}
p\left(  r\right)  \sim-k\sigma_{N}\int^{r}\left(  \frac{k^{2}}{V\left(
r\right)  }\right)  ^{N-1/2}dr. \label{pr}%
\end{equation}
To satisfy\ Eq. (\ref{uinf}), the factor $p\left(  r\right)  $ must contribute
at $r\rightarrow\infty$, i.e.,
\begin{equation}
p\left(  r\right)  \overset{r\rightarrow\infty}{\nsim}0. \label{prnot0}%
\end{equation}
Eqs. (\ref{pr}) and (\ref{prnot0}) require that $V\left(  r\right)  $ must
increase slower than or equally to $r^{1/\left(  N-1/2\right)  }$
when\ $r\rightarrow\infty$, so%
\begin{align}
\int_{a}^{\infty}\left\vert V\left(  r\right)  \right\vert \frac{1}%
{r^{\frac{1}{N-1/2}+1+\epsilon}}dr  &  =\int_{a}^{c}\left\vert V\left(
r\right)  \right\vert \frac{1}{r^{\frac{1}{N-1/2}+1+\epsilon}}dr+\int%
_{c}^{\infty}\left\vert V\left(  r\right)  \right\vert \frac{1}{r^{\frac
{1}{N-1/2}+1+\epsilon}}dr\nonumber\\
&  \leq\int_{a}^{c}\left\vert V\left(  r\right)  \right\vert \frac{1}%
{r^{\frac{1}{N-1/2}+1+\epsilon}}dr+\int_{c}^{\infty}r^{\frac{1}{N-1/2}-\delta
}\frac{1}{r^{\frac{1}{N-1/2}+1+\epsilon}}dr\nonumber\\
&  =\int_{a}^{c}\left\vert V\left(  r\right)  \right\vert \frac{1}{r^{\frac
{1}{N-1/2}+1+\epsilon}}dr+\int_{c}^{\infty}\frac{1}{r^{1+\epsilon+\delta}}dr,
\label{v4}%
\end{align}
where $\delta\geq0$. Then we have $\int_{a}^{\infty}\left\vert V\left(
r\right)  \right\vert \frac{1}{r^{B+1+\epsilon}}dr<\infty$ with $a>0$%
,\ $B=\frac{1}{N-1/2}$, and $\epsilon\sim0^{+}$, since Eq.(\ref{v4}) is finite.

$N$ equaling a nonzero finite integer gives the condition (\ref{cd4}) and
$N\rightarrow\infty$ gives the condition (\ref{Vpcd3}), respectively.
Especially, for $N\rightarrow\infty$, substituting the tortoise coordinate
(\ref{torcor5}) into Eq. (\ref{radialeq}) gives%
\end{paracol}
\begin{align}
\qquad \qquad \frac{d^{2}u_{l}\left(  r_{\ast}\right)  }{dr_{\ast}^{2}}   +\frac{\frac
{1}{4k}\left[  \frac{V^{\prime\prime}\left(  r\right)  }{V\left(  r\right)
}-\left(  \frac{V^{\prime}\left(  r\right)  }{V\left(  r\right)  }\right)
^{2}\right]  +\frac{V^{\prime}\left(  r\right)  }{2k^{2}\sqrt{\frac{V\left(
r\right)  }{k^{2}}-1}}}{\left(  \frac{V^{\prime}\left(  r\right)  }{4kV\left(
r\right)  }+\sqrt{\frac{V\left(  r\right)  }{k^{2}}-1}\right)  ^{2}}%
\frac{du_{l}\left(  r_{\ast}\right)  }{dr_{\ast}}
  +\frac{1}{\left(  \frac{V^{\prime}\left(  r\right)  }{4kV\left(  r\right)
}+\sqrt{\frac{V\left(  r\right)  }{k^{2}}-1}\right)  ^{2}}\left[  k^{2}%
-\frac{l\left(  l+1\right)  }{r^{2}}-V\left(  r\right)  \right]  u_{l}\left(
r_{\ast}\right)  =0.
\end{align}
\begin{paracol}{2}
\switchcolumn \noindent At $r\rightarrow\infty$, the coefficient of $du_{l}\left(  r_{\ast}\right)
/dr_{\ast}$ falls off faster than $1/r$ and then vanishes. At the meantime,
\begin{equation}
\frac{1}{\left(  \frac{V^{\prime}\left(  r\right)  }{4kV\left(  r\right)
}+\sqrt{\frac{V\left(  r\right)  }{k^{2}}-1}\right)  ^{2}}\left[  k^{2}%
-\frac{l\left(  l+1\right)  }{r^{2}}-V\left(  r\right)  \right]  \sim
-k^{2}-\frac{l\left(  l+1\right)  }{\left(  \frac{V\left(  r\right)  }{k^{2}%
}-1\right)  r^{2}}+\frac{V^{\prime}\left(  r\right)  /k}{2\frac{V\left(
r\right)  }{k^{2}}\sqrt{\frac{V\left(  r\right)  }{k^{2}}-1}}.
\end{equation}
Clearly,$\ \frac{l\left(  l+1\right)  }{\left(  \frac{V\left(  r\right)
}{k^{2}}-1\right)  r^{2}}-\frac{V^{\prime}\left(  r\right)  /k}{2\frac
{V\left(  r\right)  }{k^{2}}\sqrt{\frac{V\left(  r\right)  }{k^{2}}-1}}$ is a
short-range potential. This proves Eq. (\ref{shorteq5}).

\section{Scattering wave functions for long-range potential scattering
\label{phaseshift}}

In scattering theory, a general theory is established for short-range
potential scattering, in which a uniform expression of a scattering wave
function is given and all the information of scattering is embodied in a
scattering phase shift
\cite{friedrich2015scattering,pang2012relation,li2015heat}. Nevertheless,
there is no such a general treatment for long-range potential scattering. The
reason is that the large-distance behaviors for short-range potential
scattering are the same, but for long-range potential scattering are different.

In the above we show that by introducing tortoise coordinates, long-range
potential scattering can be converted to short-range potential scattering.
This allows us to develop a general theory for long-range potential scattering
just like that for short-range potential scattering. After converting
long-range potential scattering to short-range potential scattering, we can
also give a uniform expression of a scattering wave function under tortoise
coordinates and describe long-range-potential scattering by a phase shift. The
difference between large-distance asymptotic wave functions of different
potentials is reflected in tortoise coordinates.

Potentials vanishing at $r\rightarrow\infty$, which are discussed in section
\ref{mpowers}, can have scattering states. In the following we show that the
scattering wave function can be uniformly expressed in terms of tortoise
coordinates for both long-range potential scattering and short-range potential scattering.

\begin{tcolorbox}[boxrule=0pt,
  boxsep=0pt,
  colback={lightgray1},
  enhanced jigsaw,
  borderline west={3pt}{0pt}{lightgray2},
  sharp corners,
  before skip=10pt,
  after skip=10pt,
breakable,]
The radial equation with a potential satisfying the conditions (\ref{cd1}) and
(\ref{cd2}) has two linear independent solutions, $F_{l}\left(  r\right)  $
and $G_{l}\left(  r\right)  $ which satisfy the condition (\ref{radialwfrstar}%
). The large-distance asymptotic wave function can be then expressed as%
\begin{align}
u_{l}\left(  r\right)    & =C_{l}F_{l}\left(  r_{\ast}\right)  +D_{l}%
G_{l}\left(  r_{\ast}\right) \\ \nonumber
 & \overset{r\rightarrow\infty}{\sim}C_{l}e^{-i\left[  kr_{\ast}-\left(
l+1\right)  \pi/2\right]  }+D_{l}e^{i\left[  kr_{\ast}-\left(  l+1\right)
\pi/2\right]  }. \label{uexp}%
\end{align}
By introducing the phase shift%
\begin{equation}
e^{2i\delta_{l}}=\frac{D_{l}}{C_{l}},
\end{equation}
the large-distance asymptotic scattering wave function can be then written as%

\begin{equation}
u_{l}\left(  r\right)  \overset{r\rightarrow\infty}{\sim}A_{l}\sin\left(
kr_{\ast}-\frac{l\pi}{2}+\delta_{l}\right)  . \label{uscatt}%
\end{equation}
\end{tcolorbox}

Comparing with short-range potential scattering in which $u_{l}\left(
r\right)  \overset{r\rightarrow\infty}{\sim}A_{l}\sin\left(  kr-\frac{l\pi}%
{2}+\delta_{l}\right)  $, we can see that the asymptotic scattering wave
function for long-range potential scattering is just to replace coordinate $r$
with the tortoise coordinate $r_{\ast}$ in the asymptotic wave function of
short-range potential scattering.

\section{Alternative expressions of tortoise coordinates \label{Alternative}}

In this section, we give another expression of tortoise coordinates
(\ref{torcor1}) and (\ref{torcor2}) in terms of the Gauss hypergeometric function.

\subsection{The tortoise coordinate for potentials vanishing at $r\rightarrow
\infty${}}
\begin{tcolorbox}[boxrule=0pt,
  boxsep=0pt,
  colback={lightgray1},
  enhanced jigsaw,
  borderline west={3pt}{0pt}{lightgray2},
  sharp corners,
  before skip=10pt,
  after skip=10pt,
breakable,]
\textit{For potentials vanishing at }$r\rightarrow\infty$\textit{, the
tortoise coordinate (\ref{torcor1}) can be expressed as}%
\begin{equation}
r_{\ast}=\int^{r}dr\sqrt{1-\frac{V\left(  r\right)  }{k^{2}}}+\frac
{\Gamma\left(  N+1/2\right)  }{2\sqrt{\pi}\Gamma\left(  N+2\right)  }\int%
^{r}dr\,_{2}F_{1}\left(
\begin{array}
[c]{c}%
1,N+1/2\\
N+2
\end{array}
;\frac{V\left(  r\right)  }{k^{2}}\right)  \left(  \frac{V(r)}{k^{2}}\right)
^{N+1}. \label{torcor11}%
\end{equation}
\textit{Here }$_{2}F_{1}\left(
\begin{array}
[c]{c}%
a,b\\
c
\end{array}
;z\right)  =\sum_{\eta=0}^{\infty}\frac{\Gamma\left(  a+\eta\right)
\Gamma\left(  b+\eta\right)  \Gamma\left(  c\right)  }{\Gamma\left(  a\right)
\Gamma\left(  b\right)  \Gamma\left(  c+\eta\right)  \eta!}z^{\eta}$\textit{
is the Gauss hypergeometric function \cite{olver2010nist}.}
\end{tcolorbox}

\textit{Proof.} Expand $\sqrt{1-\frac{V\left(  r\right)  }{k^{2}}}$ as%
\begin{align}
\sqrt{1-\frac{V\left(  r\right)  }{k^{2}}}  &  =\sum_{\eta=0}^{\infty}\left(
-1\right)  ^{\eta}\frac{\sqrt{\pi}}{2\Gamma\left(  3/2-\eta\right)
\Gamma\left(  \eta+1\right)  }\left(  \frac{V\left(  r\right)  }{k^{2}%
}\right)  ^{\eta}\nonumber\\
&  =\sum_{\eta=0}^{N}\left(  -1\right)  ^{\eta}\frac{\sqrt{\pi}}%
{2\Gamma\left(  3/2-\eta\right)  \eta!}\left(  \frac{V\left(  r\right)
}{k^{2}}\right)  ^{\eta}\nonumber\\
&  +\sum_{\eta=0}^{\infty}\left(  -1\right)  ^{\eta+N+1}\frac{\sqrt{\pi}%
}{2\Gamma\left(  1/2-\eta-N\right)  \Gamma\left(  \eta+N+2\right)  }\left(
\frac{V\left(  r\right)  }{k^{2}}\right)  ^{\eta+N+1}. \label{expd3}%
\end{align}
Eq. (\ref{expd3}) can be rewritten as
\end{paracol}
\begin{align}
\qquad \qquad\qquad\sqrt{1-\frac{V\left(  r\right)  }{k^{2}}}    =-\sum_{\eta=0}^{N}\frac
{\Gamma\left(  \eta-1/2\right)  }{2\sqrt{\pi}\eta!}\left(  \frac{V\left(
r\right)  }{k^{2}}\right)  ^{\eta}
 -\frac{\Gamma\left(  N+1/2\right)  }{2\sqrt{\pi}\Gamma\left(  N+2\right)
}\left(  \frac{V\left(  r\right)  }{k^{2}}\right)  ^{N+1}\sum_{\eta=0}%
^{\infty}\frac{\Gamma\left(  1+\eta\right)  \Gamma\left(  N+1/2+\eta\right)
\Gamma\left(  N+2\right)  }{\Gamma\left(  N+1/2\right)  \Gamma\left(
\eta+N+2\right)  \eta!}\left(  \frac{V\left(  r\right)  }{k^{2}}\right)
^{\eta} .\label{expd4}%
\end{align}
\begin{paracol}{2}
\switchcolumn \noindent
 by use of the reflection formula of the gamma, $\Gamma\left(  -z\right)
\Gamma\left(  z+1\right)  =-\frac{\pi}{\sin\left(  \pi z\right)  }$. Comparing
with the Gauss hypergeometric function gives%
\begin{equation}
\sqrt{1-\frac{V\left(  r\right)  }{k^{2}}}=-\sum_{\eta=0}^{N}\frac
{\Gamma\left(  \eta-1/2\right)  }{2\sqrt{\pi}\eta!}\left(  \frac{V\left(
r\right)  }{k^{2}}\right)  ^{\eta}-\frac{\Gamma\left(  N+1/2\right)  }%
{2\sqrt{\pi}\Gamma\left(  N+2\right)  }\left(  \frac{V\left(  r\right)
}{k^{2}}\right)  ^{N+1}\,_{2}F_{1}\left(
\begin{array}
[c]{c}%
1,N+1/2\\
N+2
\end{array}
;\frac{V\left(  r\right)  }{k^{2}}\right)  .
\end{equation}
This proves Eq. (\ref{torcor11}).

\subsection{The tortoise coordinate for potentials nonvanishing at
$r\rightarrow\infty$}

\end{paracol}
\begin{flushright}
\begin{tcolorbox}[width=0.9\textwidth,boxrule=0pt,
  boxsep=0pt,
  colback={lightgray1},
  enhanced jigsaw,
  borderline west={3pt}{0pt}{lightgray2},
  sharp corners,
  before skip=10pt,
  after skip=10pt,
breakable,]
\textit{ For potentials nonvanishing at }$r\rightarrow\infty$\textit{, the
tortoise coordinate (\ref{torcor2}) can be expressed as}%

\begin{align}
 r_{\ast}    =\frac{1}{4k}\ln\frac{V\left(  r\right)  }{k^{2}}+\int^{r}%
dr\sqrt{\frac{V\left(  r\right)  }{k^{2}}}\sqrt{1-\frac{k^{2}}{V\left(
r\right)  }}
  +\frac{\Gamma\left(  N+1/2\right)  }{2\sqrt{\pi}\Gamma\left(  N+2\right)
}\int^{r}dr\,_{2}F_{1}\left(
\begin{array}
[c]{c}%
1,N+1/2\\
N+2
\end{array}
;\frac{k^{2}}{V\left(  r\right)  }\right)  \left(  \frac{k^{2}}{V\left(
r\right)  }\right)  ^{N+1/2}. \label{torcor22}%
\end{align}
\end{tcolorbox}
\end{flushright}
\begin{paracol}{2}
\switchcolumn 

The proof is similar to the proof of Eq. (\ref{torcor11}).

Notice that corresponding to $N\rightarrow\infty$, the constant potential is a
special case of both the above two kinds of potentials, potentials vanishing
at $r\rightarrow\infty$ corresponding to the tortoise coordinate
(\ref{torcor1}) and nonvanishing at $r\rightarrow\infty$ corresponding to the
tortoise coordinate (\ref{torcor2}).

\section{Examples: Scattering states and bound states \label{Examples}}

\subsection{The Coulomb potential}

The Coulomb potential%
\begin{equation}
V\left(  r\right)  =\frac{\alpha}{r}%
\end{equation}
corresponds to $N=1$ in the conditions (\ref{cd1}) and (\ref{cd2}). Then by
Eq. (\ref{torcor1}), we obtain the tortoise coordinate for the Coulomb
potential:%
\begin{equation}
r_{\ast}=r-\frac{\alpha}{2k^{2}}\ln r. \label{colutor}%
\end{equation}
Substituting the tortoise coordinate (\ref{colutor}) into the large-distance
asymptotics (\ref{radialwfrstar}) gives the large-distance asymptotic wave
function of the Coulomb potential%
\begin{equation}
u_{l}\left(  r_{\ast}\right)  \overset{r\rightarrow\infty}{\sim}e^{\pm
ikr_{\ast}}=\exp\left(  \pm i\left(  kr-\frac{\alpha}{2k}\ln r\right)
\right)  . \label{ucolm}%
\end{equation}

This result can also be checked by the exact solution of the Coulomb
potential,%
\begin{equation}
u_{l}\left(  r\right)  =M_{i\alpha/\left(  2k\right)  ,l+1/2}\left(
2ikr\right)  , \label{exactCp}%
\end{equation}
where $M_{\mu,\nu}\left(  z\right)  $ is the Whittaker hypergeometric
function. The large-distance asymptotics of the radial solution Eq.
(\ref{exactCp}) is%
\end{paracol}
\begin{align}
\qquad\qquad\qquad\qquad u_{l}\left(  r\right)    \overset{r\rightarrow\infty}{\sim}A_{l}%
i^{l+1}\Gamma\left(  2l+2\right)  \left[  \frac{\left(  -i\right)
^{l+1}\left(  2ik\right)  ^{-\frac{i\alpha}{2k}}}{\Gamma\left(  l-\frac
{i\alpha}{2k}+1\right)  }\exp\left(  i\left(  kr-\frac{\alpha}{2k}\ln
r\right)  \right)  \right. 
 +\left.  \frac{i^{l+1}\left(  -2ik\right)  ^{\frac{i\alpha}{2k}}}%
{\Gamma\left(  l+\frac{i\alpha}{2k}+1\right)  }\exp\left(  -i\left(
kr-\frac{\alpha}{2k}\ln r\right)  \right)  \right]  .
\end{align}
\begin{paracol}{2}
\switchcolumn \noindent
This agrees with the large-distance asymptotics given by the tortoise
coordinate, Eq. (\ref{ucolm}).

Under the tortoise coordinate (\ref{colutor}), the long-range Coulomb
potential is converted to a short-range potential. Then by Eqs. (\ref{uscatt})
and (\ref{colutor}), the scattering wave function can be represented by a
scattering phase shift:%
\begin{align}
u_{l}\left(  r\right)   &  \overset{r\rightarrow\infty}{\sim}A_{l}\sin\left(
kr_{\ast}-\frac{l\pi}{2}+\delta_{l}\right) \nonumber\\
&  =A_{l}\sin\left(  kr-\frac{\alpha}{2k}\ln r-\frac{l\pi}{2}+\delta
_{l}\right)  ,
\end{align}
where the phase shift $\delta_{l}=\Gamma\left(  l+1+\frac{i\alpha}{2k}\right)
/\Gamma\left(  l+1-\frac{i\alpha}{2k}\right)  $ is the scattering phase shift
of the Coulomb potential \cite{fl1994practical}.

\subsection{The harmonic-oscillator potential}

The harmonic-oscillator potential%
\begin{equation}
V\left(  r\right)  =\omega^{2}r^{2}%
\end{equation}
corresponds to $N=1$ in the conditions (\ref{cd3}) and (\ref{cd4}). Then by
Eq. (\ref{torcor2}), we obtain the tortoise coordinate for the
harmonic-oscillator potential:%
\begin{equation}
r_{\ast}=\frac{1}{2k}\omega r^{2}-\frac{k}{2\omega}\ln r+\frac{1}{2k}\ln\omega
r. \label{hosctor}%
\end{equation}
Substituting the tortoise coordinate (\ref{hosctor}) into the large-distance
asymptotics (\ref{uinf}) gives the large-distance asymptotic wave function of
the harmonic-oscillator potential:%
\begin{align}
u_{l}\left(  r_{\ast}\right)   &  \overset{r\rightarrow\infty}{\sim
}e^{-kr_{\ast}}\nonumber\\
&  =\exp\left(  -\left(  \frac{\omega}{2}r^{2}-\frac{k^{2}}{2\omega}\ln
r+\frac{1}{2}\ln\omega r\right)  \right)  . \label{uhoscinf}%
\end{align}

This result can also be checked by the exact solution of the
harmonic-oscillator potential,%
\begin{equation}
u_{l}\left(  r\right)  =\frac{1}{\sqrt{r}}M_{-k^{2}/\left(  4\omega\right)
,\left(  2l+1\right)  /4}\left(  \omega r^{2}\right)  . \label{uhosc}%
\end{equation}
The large-distance asymptotics of the radial solution (\ref{uhosc}) is%
\begin{equation}
u_{l}\left(  r\right)  \overset{r\rightarrow\infty}{\sim}A_{l}\omega
^{\frac{k^{2}}{4\omega}+\frac{1}{4}+\frac{l}{2}}\exp\left(  -\frac{\omega}%
{2}r^{2}+\frac{k^{2}}{2\omega}\ln r-\frac{1}{2}\ln\omega r\right)  .
\end{equation}
This agrees with the large-distance asymptotics given by the tortoise
coordinate, Eq. (\ref{uhoscinf}).

\subsection{The $1/\sqrt{r}$-potential}

The $1/\sqrt{r}$-potential%
\begin{equation}
V\left(  r\right)  =\frac{\xi}{\sqrt{r}} \label{Vrm12}%
\end{equation}
corresponds to $N=2$ in the conditions (\ref{cd1}) and (\ref{cd2}). The
tortoise coordinate of the $1/\sqrt{r}$-potential can be obtained by Eq.
(\ref{torcor1}):%
\begin{equation}
r_{\ast}=r-\frac{\xi}{k^{2}}\sqrt{r}-\frac{\xi^{2}}{8k^{4}}\ln r.
\label{toro12}%
\end{equation}
The large-distance asymptotic radial wave function of the $1/\sqrt{r}%
$-potential can be obtained by substituting the tortoise coordinate
(\ref{toro12}) into Eq. (\ref{radialwfrstar}):%
\begin{align}
u_{l}\left(  r_{\ast}\right)   &  \overset{r\rightarrow\infty}{\sim}e^{\pm
ikr_{\ast}}\nonumber\\
&  =\exp\left(  \pm i\left(  kr-\frac{\xi}{k}\sqrt{r}-\frac{\xi^{2}}{8k^{3}%
}\ln r\right)  \right)  . \label{uo12}%
\end{align}

To check this result, we calculate the large-distance asymptotes of the radial
wave function from the exact solution \cite{li2016exact}%
\begin{equation}
u_{l}\left(  r\right)  =A_{l}\left(  -2ikr\right)  ^{l+1}N\left(
4l+2,-\frac{2\xi}{\sqrt{2ik^{3}}},-\frac{\xi^{2}}{2ik^{3}},0,\sqrt
{-2ikr}\right)  \exp\left(  i\left(  kr-\frac{\xi}{k}\sqrt{r}\right)  \right)
, \label{Vrm12exact}%
\end{equation}
where $N(\alpha,\beta,\gamma,\delta,z)$ is the biconfluent Heun function
\cite{ronveaux1995heun,slavyanov2000special,li2016exact}. The large-distance
asymptotics of the exact solution (\ref{Vrm12exact}) at $r\rightarrow\infty
$\ is \cite{li2016exact}%
\begin{align}
u_{l}\left(  r\right)   &  \overset{r\rightarrow\infty}{\sim}A_{l}K_{1}\left(
4l+2,-\frac{2\xi}{\sqrt{2ik^{3}}},-\frac{\xi^{2}}{2ik^{3}},0\right)  \left(
-2ik\right)  ^{-i\alpha^{2}/\left(  8k^{3}\right)  }\exp\left(  i\left(
kr-\frac{\xi}{k}\sqrt{r}-\frac{\xi^{2}}{8k^{3}}\ln r\right)  \right)
\nonumber\\
&  +A_{l}K_{2}\left(  4l+2,-\frac{2\xi}{\sqrt{2ik^{3}}},-\frac{\xi^{2}%
}{2ik^{3}},0\right)  \left(  -2ik\right)  ^{i\alpha^{2}/\left(  8k^{3}\right)
}\exp\left(  -i\left(  kr-\frac{\xi}{k}\sqrt{r}-\frac{\xi^{2}}{8k^{3}}\ln
r\right)  \right)  ,
\end{align}
where $K_{1}\left(  4l+2,-\frac{2\xi}{\sqrt{2ik^{3}}},-\frac{\xi^{2}}{2ik^{3}%
},0\right)  $ and $K_{2}\left(  4l+2,-\frac{2\xi}{\sqrt{2ik^{3}}},-\frac
{\xi^{2}}{2ik^{3}},0\right)  $ are linear combination coefficients. This
agrees with the result obtained in virtue of the\ tortoise coordinate, Eq.
(\ref{uo12}).

Under the tortoise coordinate (\ref{toro12}), the long-range $1/\sqrt{r}%
$-potential is converted to a short-range potential. Then by Eqs.
(\ref{uscatt}) and (\ref{toro12}), the scattering wave function can be
represented by a scattering phase shift:%
\begin{align}
u_{l}\left(  r\right)   &  \overset{r\rightarrow\infty}{\sim}A_{l}\sin\left(
kr_{\ast}-\frac{l\pi}{2}+\delta_{l}\right) \nonumber\\
&  =A_{l}\sin\left(  kr-\frac{\xi}{k}\sqrt{r}-\frac{\xi^{2}}{8k^{3}}\ln
r-\frac{l}{2}\pi+\delta_{l}\right)  ,
\end{align}
where $\delta_{l}=-\arg K_{2}\left(  4l+2,\frac{2\xi}{\sqrt{2ik^{3}}}%
,\frac{\xi^{2}}{2ik^{3}},0\right)  $ is the scattering phase shift of the
$1/\sqrt{r}$-potential potential, where the definition of $K_{2}\left(
\alpha,\beta,\gamma,z\right)  $ can be found in Ref. \cite{li2016exact}.

\subsection{The $r^{2/3}$-potential}

The $r^{2/3}$-potential%
\begin{equation}
V\left(  r\right)  =\zeta r^{2/3} \label{Vr23}%
\end{equation}
corresponds to $N=2$ in the conditions (\ref{cd3}) and (\ref{cd4}). The
tortoise coordinate of the $r^{2/3}$-potential can be obtained by Eq.
(\ref{torcor2}):%
\begin{equation}
r_{\ast}=\frac{3\zeta^{1/2}}{4k}r^{4/3}-\frac{3k}{4\zeta^{1/2}}r^{2/3}%
-\frac{k^{3}}{8\zeta^{3/2}}\ln r+\frac{1}{4k}\ln\left(  \zeta r^{2/3}\right)
. \label{tor23}%
\end{equation}
The large-distance radial wave function of the $r^{2/3}$-potential can be
obtained by substituting the tortoise coordinate (\ref{tor23}) into Eq.
(\ref{uinf}):%
\begin{align}
u_{l}\left(  r_{\ast}\right)   &  \overset{r\rightarrow\infty}{\sim
}e^{-kr_{\ast}}\nonumber\\
&  =\exp\left(  -\left(  \frac{3\zeta^{1/2}}{4}r^{4/3}-\frac{3k^{2}}%
{4\zeta^{1/2}}r^{2/3}-\frac{k^{4}}{8\zeta^{3/2}}\ln r+\frac{1}{4}\ln\left(
\zeta r^{2/3}\right)  \right)  \right)  . \label{u23inf}%
\end{align}

To check the above result, we calculate the large-distance asymptotics of the
radial wave function from the exact solution
\end{paracol}
\begin{align}
\qquad\qquad \qquad \qquad \qquad  u_{l}\left(  r\right)    =A_{l}\exp\left(  -\frac{3}{4}\zeta^{1/2}%
r^{4/3}+\frac{3k^{2}}{4\zeta^{1/2}}r^{2/3}\right)  \left(  \frac{\sqrt{6}}%
{2}\zeta^{1/4}\right)  ^{3\left(  l+1\right)  /2}r^{l+1}
 N\left(  3l+\frac{3}{2},-\frac{\sqrt{6}k^{2}}{2\zeta^{3/4}}%
,\frac{3k^{4}}{8\zeta^{3/2}},0,\frac{\sqrt{6}}{2}\zeta^{1/4}r^{2/3}\right)  .
\label{u23}%
\end{align}
\begin{paracol}{2}
\switchcolumn \noindent The large-distance asymptotics of the exact solution (\ref{u23}) at
$r\rightarrow\infty$\ is%
\begin{equation}
u_{l}\left(  r\right)  \overset{r\rightarrow\infty}{\sim}\exp\left(  -\frac
{3}{4}\zeta^{1/2}r^{4/3}+\frac{3k^{2}}{4\zeta^{1/2}}r^{2/3}+\frac{k^{4}%
}{8\zeta^{3/2}}\ln r+\frac{1}{6}\ln r\right)  .\nonumber
\end{equation}
This agrees with the result obtained in virtue of the\ tortoise coordinate,
Eq. (\ref{u23inf}).

\subsection{The $1/r^{3/2}$-potential}

The tortoise coordinate of the $1/r^{3/2}$-potential%
\begin{equation}
V\left(  r\right)  =\frac{\xi}{r^{3/2}}%
\end{equation}
by Eq. (\ref{torcor1}) is
\begin{equation}
r_{\ast}=r, \label{toro32}%
\end{equation}
i.e., the tortoise coordinate of the $1/r^{3/2}$-potential is the radial
coordinate $r$ itself. This means that the $1/r^{3/2}$-potential is in fact a
short-range potential. Then the large-distance asymptotic radial wave function
of the $1/r^{3/2}$-potential is just the large-distance asymptotic radial wave
function of all short-range potentials:
\begin{align}
u_{l}\left(  r_{\ast}\right)   &  \overset{r\rightarrow\infty}{\sim}e^{\pm
ikr_{\ast}}\nonumber\\
&  =e^{\pm ikr}. \label{uo32}%
\end{align}

The large-distance asymptotic radial wave function of the $1/r^{3/2}%
$-potential can be obtained from the\ exact solution%
\begin{equation}
u_{l}\left(  r\right)  =A_{l}e^{ikr}\left(  -2ikr\right)  ^{l+1}N\left(
4l+2,0,0,-\frac{8\xi}{\sqrt{-2ik}},-\sqrt{-2ikr}\right)  . \label{esVr32}%
\end{equation}
The large-distance asymptotics of the exact solution\ (\ref{esVr32}) at
$r\rightarrow\infty$ is%
\begin{equation}
u_{l}\left(  r\right)  \overset{r\rightarrow\infty}{\sim}A_{l}K_{1}\left(
4l+2,0,0,-\frac{8\xi}{\sqrt{-2ik}}\right)  e^{ikr}+A_{l}K_{2}\left(
4l+2,0,0,-\frac{8\xi}{\sqrt{-2ik}}\right)  e^{-ikr}.
\end{equation}
This agrees with the result given by the\ tortoise coordinate, Eq. (\ref{uo32}).

The $1/r^{3/2}$-potential is indeed a short-range potential. By Eqs.
(\ref{uscatt}) and (\ref{toro32}), the scattering wave function can be
represented by a scattering phase shift:%
\begin{align}
u_{l}\left(  r\right)   &  \overset{r\rightarrow\infty}{\sim}A_{l}\sin\left(
kr_{\ast}-\frac{l\pi}{2}+\delta_{l}\right) \nonumber\\
&  =A_{l}\sin\left(  kr-\frac{l\pi}{2}+\delta_{l}\right)  ,
\end{align}
where $\delta_{l}=-\arg K_{2}\left(  4l+2,0,0,\frac{8\xi}{\sqrt{-2ik}}\right)
$ is the scattering phase shift of the $1/r^{3/2}$-potential.

\subsection{The $r^{6}$-potential}

The $r^{6}$-potential%
\begin{equation}
V\left(  r\right)  =\zeta r^{6}%
\end{equation}
has only bound states. The tortoise coordinate of the $r^{6}$-potential by Eq.
(\ref{torcor1}) reads%
\begin{equation}
r_{\ast}=\frac{1}{4k}\zeta^{1/2}r^{4}+\frac{1}{4k}\ln\left(  \zeta
r^{6}\right)  . \label{tor6}%
\end{equation}
The large-distance asymptotic radial wave function of the $r^{6}$-potential
can be obtained by substituting the tortoise coordinate (\ref{tor6}) into Eq.
(\ref{uinf}):%
\begin{align}
u_{l}\left(  r_{\ast}\right)   &  \overset{r\rightarrow\infty}{\sim
}e^{-kr_{\ast}}\nonumber\\
&  =\exp\left(  -\frac{1}{4}\zeta^{1/2}r^{4}-\frac{1}{4}\ln\left(  \zeta
r^{6}\right)  \right)  . \label{u6inf}%
\end{align}

To check the above result, we calculate the large-distance asymptotic radial
wave function from the exact solution of the $r^{6}$-potential
\begin{equation}
u_{l}\left(  r\right)  =A_{l}\left(  \frac{\zeta^{1/2}}{2}\right)  ^{\left(
l+1\right)  /4}\exp\left(  -\frac{1}{4}\zeta^{1/2}r^{4}\right)  r^{l+1}%
N\left(  l+\frac{1}{2},0,0,\frac{k^{2}}{\sqrt{2}\zeta^{1/4}},\frac{1}{\sqrt
{2}}\zeta^{1/4}r^{2}\right)  . \label{u6}%
\end{equation}
The large-distance asymptotics of the exact solution (\ref{u6}) at
$r\rightarrow\infty$\ is%
\begin{equation}
u_{l}\left(  r\right)  \overset{r\rightarrow\infty}{\sim}A_{l}\exp\left(
-\frac{1}{4}\zeta^{1/2}r^{4}-\frac{3}{2}\ln r\right)  .
\end{equation}
This agrees with the result given by the\ tortoise coordinate, Eq.
(\ref{u6inf}).

\section{The duality: tortoise coordinates and asymmetric wave functions
\label{duality}}

\subsection{The duality relation}

In section \ref{Classification}, we classify potentials in terms of tortoise
coordinates. Newton discovered a duality in classical mechanics, called the
Newton-Hooke duality \cite{chandrasekhar1995newton}. Kasner and Arnol'd
generalized this duality to arbitrary power potentials in classical mechanics
\cite{arnold2013mathematical,arnold1990huygens,needham1998visual,needham1993newton,hall2000planetary}%
.\ Ref. \cite{li2021duality} gives the general result of this duality,
including the duality of arbitrary potentials in classical and quantum
mechanics and in scalar fields. In the following, we discuss the relation
between the duality and the classification of potentials.

Consider two central potentials $V\left(  r\right)  $ and $U\left(
\rho\right)  $. Their radial equations are
\begin{align}
\frac{d^{2}u_{l}\left(  r\right)  }{dr^{2}}+\left[  k^{2}-\frac{l\left(
l+1\right)  }{r^{2}}-V\left(  r\right)  \right]  u_{l}\left(  r\right)   &
=0,\\
\frac{d^{2}v_{\ell}\left(  \rho\right)  }{d\rho^{2}}+\left[  \kappa^{2}%
-\frac{\ell\left(  \ell+1\right)  }{\rho^{2}}-U\left(  \rho\right)  \right]
v_{\ell}\left(  \rho\right)   &  =0.
\end{align}
Suppose the potentials $V\left(  r\right)  $ and $U\left(  \rho\right)  $ can
be expanded as a general polynomial which is a polynomial with arbitrary real
number powers:%
\begin{align}
V\left(  r\right)   &  =\xi r^{a}+\sum_{b_{n}}\mu_{n}r^{b_{n}},\text{ \ }%
b_{n}<a,\label{Vpot}\\
U\left(  \rho\right)   &  =\zeta\rho^{A}+\sum_{B_{n}}\lambda_{n}\rho^{B_{n}%
},\text{ \ }B_{n}<A. \label{Upot}%
\end{align}
According to the duality\ relation between $V\left(  r\right)  $ and $U\left(
\rho\right)  $ given in Ref. \cite{li2021duality}, the duality relation
between the expansions (\ref{Vpot}) and (\ref{Upot}) is
\begin{align}
\frac{a+2}{2}  &  =\frac{2}{A+2},\label{dual1}\\
\sqrt{\frac{2}{a+2}}\left(  b_{n}+2\right)   &  =\sqrt{\frac{2}{A+2}}\left(
B_{n}+2\right)  . \label{dual2}%
\end{align}
The wave functions of these two systems can be transformed into each other by
the duality transform%
\begin{align}
r  &  \rightarrow\rho^{\left(  A+2\right)  /2},\label{coorTpower}\\
u_{l}\left(  r\right)   &  \rightarrow\rho^{A/4}v_{\ell}\left(  \rho\right)  ,
\label{wTpower}%
\end{align}
and%
\begin{align}
k^{2}  &  \rightarrow-\zeta\left(  \frac{2}{A+2}\right)  ^{2},\label{coup}\\
\xi &  \rightarrow-{\kappa}^{2}\left(  \frac{2}{A+2}\right)  ^{2},
\label{en}%
\end{align}%
\begin{equation}
l+\frac{1}{2}\rightarrow\frac{2}{A+2}\left(  \ell+\frac{1}{2}\right)  .
\label{ltrans}%
\end{equation}

By the above duality relation, we obtain the following duality relation
between tortoise coordinates and asymptotic wave functions.

\begin{tcolorbox}[boxrule=0pt,
  boxsep=0pt,
  colback={lightgray1},
  enhanced jigsaw,
  borderline west={3pt}{0pt}{lightgray2},
  sharp corners,
  before skip=10pt,
  after skip=10pt,
breakable,]
\textit{If the central potential }$V\left(  r\right)  $\textit{\ satisfies the
conditions (\ref{cd1}) and (\ref{cd2}) at infinity, then its dual potential
}$U\left(  \rho\right)  $\textit{ satisfies the conditions (\ref{cd3}) and
(\ref{cd4}) at infinity. The dual potentials }$V\left(  r\right)  $\textit{
and }$U\left(  \rho\right)  $\textit{ correspond to the same value of the
positive integer }$N$\textit{, and their tortoise coordinates and asymptotic
wave functions are related by the duality transforms (\ref{coorTpower}%
)-(\ref{ltrans}).}
\end{tcolorbox}

\textit{Proof.} (1) Suppose that at $r\rightarrow\infty$, the potential
$V\left(  r\right)  $ satisfies the conditions (\ref{cd1}) and (\ref{cd2}),
i.e., $V\left(  r\right)  $ falls off faster than $\frac{1}{r^{1/\left(
N+1\right)  }}$ and slower than or equally to $\frac{1}{r^{1/N}}$; or,
equivalently,%
\begin{equation}
-\frac{1}{N}\leq a<-\frac{1}{N+1},\text{ \ }N\in Z^{+}. \label{aeq}%
\end{equation}

By the duality relation (\ref{dual1}) and the expansion of the dual potentials
$V\left(  r\right)  $ and $U\left(  \rho\right)  $, we have
\begin{equation}
A=-\frac{2a}{2+a}. \label{Aanda}%
\end{equation}
Substituting Eq. (\ref{Aanda}) into Eq. (\ref{aeq}) gives
\begin{equation}
\frac{2}{2N+1}<A\leq\frac{2}{2N-1}.
\end{equation}
Moreover, by the duality relation (\ref{dual2}) we have
\begin{equation}
B_{n}<A.
\end{equation}
This proves that $U\left(  \rho\right)  $ satisfies the conditions (\ref{cd3})
and (\ref{cd4}), i.e., $U\left(  \rho\right)  $ falls off faster than
$r^{\frac{2}{2N+1}}$ and and slower than or equally to $r^{\frac{2}{2N-1}}$.
Therefore, the tortoise coordinate and asymptotic wave function of $U\left(
\rho\right)  $ are given by (\ref{torcor2}) and (\ref{uinf}).

(2) Next, we show that the asymptotic wave function of $U\left(  \rho\right)
$ can be obtained by performing the duality transforms (\ref{coorTpower}%
)-(\ref{en}).

Because at $r\rightarrow\infty$, $V\left(  r\right)  $ falls off as $\frac
{1}{r^{\frac{1}{N+1}}}<\frac{1}{r^{a}}\leq\frac{1}{r^{\frac{1}{N}}}$, by Eqs.
(\ref{torcor1}) and (\ref{radialwfrstar}), the asymptotic wave function of
$V\left(  r\right)  $ at $r\rightarrow\infty$ is%
\begin{equation}
u_{l}\left(  r\right)  \overset{r\rightarrow\infty}{\sim}\exp\left(  \pm
ik\left[  r-\sum_{\eta=1}^{N}\frac{\Gamma\left(  \eta-\frac{1}{2}\right)
}{2\sqrt{\pi}\eta!k^{2\eta}}\int V\left(  r\right)  ^{\eta}dr\right]  \right)
.
\end{equation}
Taking only the leading contribution into account gives
\begin{equation}
u_{l}\left(  r\right)  \overset{r\rightarrow\infty}{\sim}\exp\left(  \pm
ik\left[  r-\sum_{\eta=1}^{N}\frac{\Gamma\left(  \eta-\frac{1}{2}\right)
}{2\sqrt{\pi}\eta!k^{2\eta}}\int\left(  \frac{\xi}{r^{a}}\right)  ^{\eta
}dr\right]  \right)  . \label{Vinf}%
\end{equation}
Substituting the duality relations (\ref{coorTpower})-(\ref{en}) into Eq.
(\ref{Vinf}) gives
\begin{align}
&  \rho^{A/4}v_{\ell}\left(  \rho\right)  \overset{\rho\rightarrow\infty
}{\sim}\exp\left(  \pm i\sqrt{-\zeta\left(  \frac{2}{A+2}\right)  ^{2}}\left[
\rho^{\left(  A+2\right)  /2}\right.  \right. \nonumber\\
&  -\left.  \left.  \sum_{\eta=1}^{N}\frac{\Gamma\left(  \eta-\frac{1}%
{2}\right)  }{2\sqrt{\pi}\eta!\left(  -\zeta\left(  \frac{2}{A+2}\right)
^{2}\right)  ^{\eta}}\int\left(  \frac{-\mathcal{\kappa}^{2}\left(  \frac
{2}{A+2}\right)  ^{2}}{\left(  \rho^{\left(  A+2\right)  /2}\right)
^{\frac{2A}{A+2}}}\right)  ^{\eta}d\rho^{\left(  A+2\right)  /2}\right]
\right)  . \label{vrho}%
\end{align}
Note that by Eq. (\ref{coorTpower}) we can see that $\rho\rightarrow\infty$
when $r\rightarrow\infty$.

Rewriting Eq. (\ref{vrho}) as
\begin{equation}
v_{\ell}\left(  \rho\right)  \overset{\rho\rightarrow\infty}{\sim}\frac
{1}{\left[  U\left(  \rho\right)  \right]  ^{1/4}}\exp\left(  \mp\left[
\int\sqrt{U\left(  \rho\right)  }d\rho-\sum_{\eta=1}^{N}\frac{\Gamma\left(
\eta-\frac{1}{2}\right)  \mathcal{\kappa}^{2\eta}}{2\sqrt{\pi}\eta!}\int%
\frac{1}{\left[  U\left(  \rho\right)  \right]  ^{\left(  2\eta-1\right)  /2}%
}d\rho\right]  \right)  .
\end{equation}
For potentials satisfying $V\left(  r\right)  \overset{r\rightarrow
\infty}{\sim}r^{\beta}$ ($\beta>0$), their asymptotic wave function should
tend to zero.

The asymptotic wave function of the potential satisfying $V\left(  r\right)
\overset{r\rightarrow\infty}{\sim}r^{\beta}$ ($\beta>0$) should vanish at
$r\rightarrow\infty$, so we choose minus sign:%
\begin{equation}
v_{\ell}\left(  \rho\right)  \overset{\rho\rightarrow\infty}{\sim}\frac
{1}{\left[  U\left(  \rho\right)  \right]  ^{1/4}}\exp\left(  -\int%
\sqrt{U\left(  \rho\right)  }d\rho+\sum_{\eta=1}^{M}\frac{\Gamma\left(
\eta-\frac{1}{2}\right)  \mathcal{\kappa}^{2\eta}}{2\sqrt{\pi}\eta!}\int%
\frac{1}{\left[  U\left(  \rho\right)  \right]  ^{\left(  2\eta-1\right)  /2}%
}d\rho\right)  .
\end{equation}
This is consistent with the asymptotics given by Eqs. (\ref{torcor2}) and
(\ref{uinf}).

\subsection{Classification of potentials}

In section \ref{Classification}, we classify potentials in terms of tortoise
coordinates. In this section, we show the relation between\ the classification
and the duality.

For convenience, we consider the power potential. Since the long-distance
behavior of potentials can be analyzed by their power expansion, the
conclusion drawn from the power potential is not limited to the power potential.

According to the duality relation (\ref{dual1}),\ for power potential
$V\left(  r\right)  \sim r^{a}$, we have the following conclusion.

\begin{tcolorbox}[boxrule=0pt,
  boxsep=0pt,
  colback={lightgray1},
  enhanced jigsaw,
  borderline west={3pt}{0pt}{lightgray2},
  sharp corners,
  before skip=10pt,
  after skip=10pt,
breakable,]
Case 1\textit{. The dual potentials with the powers }%
\begin{equation}
-1\leq a\leq0 \label{cls11}%
\end{equation}
\textit{are the potentials with the powers}%
\begin{equation}
0\leq a\leq2. \label{cls12}%
\end{equation}
\textit{Especially, }$V\left(  r\right)  \sim r^{0}$\textit{, i.e., }%
$a=0$\textit{, is self-dual. }

Case 2\textit{. The dual potentials with the powers}%
\begin{equation}
-2<a<-1 \label{cls21}%
\end{equation}
\textit{are the potentials with the powers}%
\begin{equation}
a>2. \label{cls22}%
\end{equation}

Case 3\textit{. The dual potentials with the powers}%
\begin{equation}
a<-2 \label{cls31}%
\end{equation}
\textit{are the potentials with the powers}%
\begin{equation}
a<-2.
\end{equation}
\end{tcolorbox}

In case 1, the potentials satisfying (\ref{cls11}) are long-range potentials,
and their dual potentials, satisfying (\ref{cls12}), are also long-range potential.

It is worth noting that in case 2, the potential satisfying (\ref{cls21}) is a
short-range potential, because their tortoise coordinates are the same as
those satisfying (\ref{cls31}). However, from the viewpoint of duality, the
dual potential of the potentials satisfying (\ref{cls21}) is also a positive
power long-range potential, but the dual potential of the short-range
\ potential satisfying (\ref{cls31}), which satisfies (\ref{cls31}), is still
a short-range potential. That is, the short-range potential satisfies
(\ref{cls21}) is somewhat special.

The above classification can also be used to discuss the existence of bound states.

If two potentials are dual, their wave functions can be obtained from each
other through the dual transform. A positive-power potential has only bound
states. After the dual transform, the bound-state wave function is still a
bound-state wave function. Therefore, the dual potential of a positive-power
potential must have bound states, although the dual potential can also have
scattering states at the same time.

More concretely, the positive-power potential $U\left(  r\right)  =\lambda
r^{a}$ with $\lambda>0$ and $a>0$ has only bound states. Its dual potential
$V\left(  \rho\right)  =\eta\rho^{b}$ whose $\eta<0$ and $-2<b<0$ must also
has bound-states.\ The duality transform transforms the bound state of
$U\left(  r\right)  $ to the bound state of its dual potential $V\left(
\rho\right)  $. But $V\left(  \rho\right)  $ also has scattering states
besides bound states. The scattering state of $V\left(  \rho\right)  $ is
dually related to the scattering state of $U\left(  r\right)  =\lambda r^{a}$
with $\lambda<0$ and $a>0$, which, however, is not lower bounded for
$\lambda<0$.

Detailed discussions on the existence condition of bound states can be found
in Refs \cite{Brau2004Necessary,Brau2004Sufficient,Calogero1965Upper}.

\section{Conclusion \label{Conclusion}}

Inspired by general relativity, we suggest a general treatment for long-range
potential scattering by introducing tortoise coordinates. While in common
scattering theory, only the short-range potential can be treated generally.

The key treatment in our scheme is to introduce the tortoise coordinate. The
tortoise coordinate in general relativity is introduced to convert a curved
spacetime to a partially conformally flat spacetime. In our scheme, the
tortoise coordinate is introduced to convert a long-range potential to a
short-range potential.

Starting from the tortoise coordinate, we suggest a classification scheme for
potentials. Newton and Euler classify functions in virtue of their asymptotics
\cite{blanton2012introduction}. The asymptotic behavior of wave functions is
reflected in the tortoise coordinate. In the paper, we classify potentials by
the corresponding tortoise coordinates. Moreover, we also show a relation
between classification of potentials and a duality in quantum mechanics.

This classification scheme is indeed a classification of the Schr\"{o}dinger
operator. The Schr\"{o}dinger operator is one of the Laplacian type operator
which consists of a Laplacian operator and a potential function. Replacing the
Laplacian operator with the Laplace-Beltrami operator, a Laplacian operator in
curved space, we can use a similar approach to classify spacetime manifolds on
which the Laplace-Beltrami operator is defined.

By tortoise coordinates, a problem of long-range potentials is converted to a
problem of short-range potentials. A short-range potential scattering can be
fully described by the scattering phase shift
\cite{liu2014scattering,li2016scattering}. The scattering phase shift can be
expressed by tortoise coordinates in gravity related scattering problems,
which are the long-range scattering
\cite{li2018scalar,li2019scattering,li2021scalar}. This inspires us to
describe long-range scattering in virtue of the tortoise coordinate by a phase
shift. Therefore, the method for the calculation of scattering phase shifts
for short-range potential scattering can be then applied to long-range
potential scattering.

Furthermore, the method suggested in the present paper can also be applied to
scattering in curved space. We will also consider the application of the
tortoise coordinate in gauge field theory.

\bigskip
\bigskip

\noindent
\textbf{ACKNOWLEDGMENTS}

\noindent
We are very indebted to Dr G. Zeitrauman for his encouragement. This work is supported in part by Special Funds for theoretical physics Research Program of the NSFC under Grant No. 11947124, and NSFC under Grant Nos. 11575125 and 11675119.

\nolinenumbers

\reftitle{References}




\providecommand{\href}[2]{#2}\begingroup\raggedright\endgroup

\end{paracol}
%


\end{document}